\documentclass[reprint,amsmath,amssymb,aps, showpacs]{revtex4-2}

\usepackage{graphicx}
\usepackage{dcolumn}
\usepackage{bm}
\usepackage{mathtools}
\usepackage{hyperref}
\usepackage{xcolor}
\usepackage{todonotes}
\usepackage{pgfplots}
\hypersetup{
    colorlinks,
    linkcolor={red!50!black},
    citecolor={blue!50!black},
    urlcolor={blue!80!black}
}
\definecolor{seabornblue}{rgb}{0.2980392156862745, 0.4470588235294118, 0.6901960784313725}
\definecolor{seaborngreen}{rgb}{0.3333333333333333, 0.6588235294117647, 0.40784313725490196}
\definecolor{seabornred}{rgb}{0.7686274509803922, 0.3058823529411765, 0.3215686274509804}
\definecolor{seabornpurple}{rgb}{0.5058823529411764, 0.4470588235294118, 0.6980392156862745}
\definecolor{seabornsand}{rgb}{0.8, 0.7254901960784313, 0.4549019607843137}
\definecolor{seaborncyan}{rgb}{0.39215686274509803, 0.7098039215686275, 0.803921568627451}
\definecolor{seabornorange}{rgb}{0.958, 0.476, 0.206}

\DeclarePairedDelimiter\abs{\lvert}{\rvert}
\DeclarePairedDelimiter\norm{\lVert}{\rVert}

\newcommand{\uz}[1]{{#1}}
\newcommand{\vg}[1]{{#1}}
\newcommand{\rev}[1]{{#1}}
\newcommand{\errata}[1]{{\color{seabornblue}#1}}

\begin{document}

\preprint{APS/123-QED}

\title{Time-harmonic waves in Korteweg and nematic-Korteweg fluids}
\thanks{}%

\author{Patrick E.~Farrell}
\email{patrick.farrell@maths.ox.ac.uk}
\affiliation{Mathematical Institute, University of Oxford, Oxford, United Kingdom and \\ Mathematical Institute, Faculty of Mathematics and Physics, Charles University, Czechia.}
\author{Umberto Zerbinati}%
 \email{zerbinati@maths.ox.ac.uk}
\affiliation{Mathematical Institute, University of Oxford, Oxford, United Kingdom.}
\date{\today}
\begin{abstract}
We derive the Helmholtz--Korteweg equation, which models acoustic waves in Korteweg fluids.
We further derive a nematic variant of the Helmholtz--Korteweg equation, which incorporates an additional orientational term in the stress tensor.
Its dispersion relation coincides with that arising in Virga's analysis of the Euler--Korteweg equations, which we 
extend to consider imaginary wave numbers and the effect of boundary conditions.
In particular, our extensions allow us to analyze the effect of nematic orientation on the penetration depth of evanescent plane waves, and on the scattering of sound waves by obstacles.
Furthermore, we make new, experimentally-verifiable predictions for the effect of boundary conditions for a modification of the Mullen--L\"{u}thi--Stephen experiment, and for the scattering of acoustic waves in nematic-Korteweg fluids by a circular obstacle.
\end{abstract}
\maketitle

A Korteweg fluid is one that incorporates a term in the stress tensor that depends on density gradients~\cite{Korteweg}. For example, an inviscid Korteweg fluid has stress tensor
\begin{equation}
    \underline{\underline{\sigma}}^{(K)} = p\underline{\underline{I}} - u_1\rho \left(\nabla\rho \otimes \nabla \rho \right),\label{eq:Korteweg}
\end{equation}
where $\underline{\underline{\sigma}}^{(K)}$ is the stress tensor, $p$ is the pressure, and $u_1 > 0$ is a material constant. For a van der Waals fluid, the pressure can be expressed as
\begin{equation}
    p = \rho c_0^2 - \rho\nabla\cdot \left[\rho u_1 \nabla \rho \right],\label{eq:pressureKorteweg}
\end{equation}
where $c_0 > 0$ is the isotropic speed of sound.
The Euler--Korteweg system is the combination of the usual Euler equations with the constitutive relation \eqref{eq:Korteweg}.

At first, Korteweg's theory had little application, as remarked upon in Truesdell et al.~\cite{TruesdellNoll}:
\begin{quote}
   Korteweg, [\dots] did not feel the second requirement of modern work, namely, to get different and new results as well.
   His theory seems not to have been taken up by any later writer.
\end{quote}
Later authors have found many applications and new results based on Korteweg's theory.
While the Korteweg term is often neglected for rarefied gases and nearly incompressible fluids, the
Korteweg stress is important when we seek to model the behavior of fluids near liquid-vapor interfaces, where the
density gradient is large.
For example, Korteweg fluids have been used to model phenomena such as capillary waves and gas-saturated magma melts \cite{LauroMagma,BenzoniPlanarWaves}.
Furthermore, Korteweg's theory has been the starting point for the development of nematoacoustic models \cite{VirgaNematoacoustics}, which are used to describe the propagation of sound waves in nematic liquid crystals.
In fact, nematic liquid crystals can be regarded as Korteweg fluids on an acoustic length scale, where the stress tensor is augmented with a term related to the nematic director field $\vec{n} \in \mathbb{S}^2$
\begin{equation}
    \label{eq:NematicKorteweg}
    \underline{\underline{\sigma}}^{(V)} = p\underline{\underline{I}} - u_1\rho \left(\nabla\rho \otimes \nabla \rho \right) - u_2 \errata{\rho}\left(\nabla \rho \cdot \vec{n} \right)\nabla \rho \otimes \vec{n}, 
\end{equation}
where $u_2 > 0$ is a material constant related to the nematic liquid crystal and the pressure is now given by 
\begin{equation}
    \label{eq:pressureNematic}
    p = \rho c_0^2 - \rho \nabla\cdot \left[\rho \left(u_1\nabla \rho + u_2 (\nabla\rho \cdot \vec{n})\vec{n} \right) \right].
\end{equation}
\rev{While general second-grade materials can convey internal torques by a hyperstress, in Korteweg fluids this vanishes identically~\cite{VirgaNematoacoustics}.}

\rev{
    The state variables for a Korteweg fluid are its density $\rho$, its gradient $\nabla \rho$, and (in the context of nematic-Korteweg fluids) also the nematic director $\vec{n}$. Under the assumption that, given an increasing function $\sigma_0(\rho)$, the Helmholtz free energy density of the fluid is given by
    \begin{equation}
        W(\rho, \nabla\rho, \vec{n}) = \sigma_0(\rho) + \frac{1}{2}u_1 \norm{\nabla \rho}^2 + \frac{1}{2}u_2 (\nabla \rho \cdot \vec{n})^2,
    \end{equation}
    the Korteweg and nematic-Korteweg stress tensors can be derived from a hyperelastic framework \cite{Capriz}.
}
A more classical derivation of Korteweg's model can be found in Toupin \cite{Toupin1,Toupin2}, where continuum equations are derived from a classical principle of virtual work.
The Korteweg equations can also be derived from a non-standard variational principle \cite{FriedGurtin} and from the kinetic theory of gases \cite{Giovangigli}.

The continuum thermodynamics of Korteweg fluids has been extensively studied by many authors, leading to the development of new concepts such as interstitial work, balance of self-equilibrated forces, and multipolarity \cite{DunnSerrin,AndersonEtAll,MehrabadiEtAll}.

Korteweg fluids have also been extensively studied from a mathematical point of view.
For example, under certain assumptions, the Euler--Korteweg equations can be formulated as a Hamiltonian system \cite{BenzoniGavageHamiltonian,BenzoniCIRM}.
Furthermore, the well-posedness of the Euler--Korteweg system has been studied in detail \cite{BenzoniGavageWellposedness,MurataShibata,Tsuda}.
Lastly, the Euler--Korteweg system has also been studied from a gradient flow perspective \cite{TzavarasEtAll}.

In this work we consider the propagation of time-harmonic pressure waves in Korteweg and particularly nematic-Korteweg fluids.
Acoustic waves in such fluids can exhibit anisotropic phenomena: the speed of sound depends on the orientation of the nematic director. Conversely, ultrasonic waves can change the nematic field, and thereby change the observed color of liquid crystals, \vg{under appropriate conditions}. Moreover, the subject is somewhat controversial in the literature, with no consensus on the physical mechanisms behind these phenomena~\cite{VirgaSonnet}. 

In this work we neglect effects on timescales longer than the acoustic timescale. The long-time asymptotic behavior of waves in Korteweg fluids has been studied by Benzoni-Gavage and coworkers~\cite{BenzoniPlanarWaves,BenzoniGavageLongAsymptotics}.

\section{Helmholtz--Korteweg equation}
Following the well-known derivation of the wave equation from Euler's equations, we begin considering the continuity equation and the balance law of linear momentum \vg{in the absence of external body forces}, i.e.
\begin{align}
    \partial_t \rho + \nabla\cdot (\rho \vec{v}) = 0, \qquad \rho \left[\partial_t\vec{v}+(\underline{\underline{\nabla \vec{v}}})\vec{v}\right]= -(\nabla \cdot \underline{\underline{\sigma}})\;, \label{eq:balanceLaw}
\end{align}
where $\vec{v}(\vec{x},t)$ is the fluid velocity and $\underline{\underline{\sigma}}$ is the Cauchy stress tensor.

We are interested in disturbances in the density field of the form $\rho(\vec{x},t) = \rho_0 \left(1+s(\vec{x},t)\right)$, where $s(\vec{x},t)$ is a time-harmonic condensation, i.e.
\begin{equation}
    s(\vec{x},t) = \Re\left[S(\vec{x})e^{-i\omega t}\right],\label{eq:timeHarmonic}
\end{equation}
with $\omega$ being the frequency of the disturbances.
Furthermore, we will assume that the condensation is a small perturbation of the density field, i.e.~$\abs{s(\vec{x},t)}=\mathcal{O}(\varepsilon)$, with $\varepsilon\ll 1$.
Lastly, we will assume that the velocity field is a small perturbation around the stationary regime, i.e.~$\norm{\vec{v}(\vec{x},t)} = \mathcal{O}(\varepsilon)$.

Under these assumptions, we can rewrite \eqref{eq:balanceLaw} as
\begin{align}
    \rho_0 \left[\partial_t s + \nabla\cdot \vec{v} + \mathcal{O}(\varepsilon^2)\right] = 0, \nonumber \\
    \partial_t\vec{v}+\mathcal{O}(\varepsilon^2) = -\rho^{-1}(\nabla \cdot \underline{\underline{\sigma}})\;.\nonumber
\end{align}
Neglecting terms of order $\mathcal{O}(\varepsilon^2)$ and observing that since $\abs{s(\vec{x}, t)}\ll 1$ we have $\rho^{-1}\approx \rho_0^{-1}$, we end up with the linearised balance law, i.e.
\begin{align}
    \rho_0\left[\partial_t s + \nabla\cdot \vec{v} \right]= 0, \qquad \partial_t\vec{v} = -\rho_0^{-1}(\nabla \cdot \underline{\underline{\sigma}})\;.\label{eq:linearisedBalanceLaw}
\end{align}
Taking the time derivative of the continuity equation and substituting for $\partial_t \vec{v}$ yields a general wave equation, i.e.
\begin{equation}
    \rho_0 \partial_t^2 s - \nabla\cdot \left(\nabla \cdot \underline{\underline{\sigma}}\right) = 0. \label{eq.generalWave}
\end{equation}
Substituting the ansatz \eqref{eq:timeHarmonic} in the general wave equation \eqref{eq.generalWave} yields
\begin{equation}
    \Re\left[-\rho_0\omega^2S(\vec{x})e^{-i\omega t}\right] = -\Re\left[\nabla\cdot \left(\nabla \cdot \underline{\underline{\sigma}}\right)\right].\label{eq:timeHarmonicGenWave}
\end{equation}
We first consider Korteweg fluids. Neglecting terms of higher order in the Korteweg constitutive relation~\eqref{eq:Korteweg} yields
\begin{equation}
    \nabla\cdot \underline{\underline{\sigma}} \approx \nabla p = c_0^2 \nabla \rho - \nabla\left[\rho \nabla\cdot (\rho u_1\nabla\rho)\right],
\end{equation}
i.e.~we consider a purely spherical response.
Using the time-harmonic ansatz \eqref{eq:timeHarmonic}, we obtain
\begin{equation}
   \Re\left[\nabla\cdot \underline{\underline{\sigma}}\right] \approx \Re\left[c_0^2 \rho_0 \nabla S(\vec{x})e^{-i\omega t} - \rho_0^3 u_1 \nabla\left(\Delta S(\vec{x})e^{-i\omega t}\right)\right].\label{eq:KortewegTimeHarmonic}
\end{equation}
We can now substitute \eqref{eq:KortewegTimeHarmonic} in \eqref{eq:timeHarmonicGenWave} and divide by $\rho_0 e^{-\omega t}$. Assuming $S(\vec{x})$ is analytic, we can extend the equation just derived to the whole complex plane and obtain the Helmholtz--Korteweg equation, i.e.
\begin{equation}
    -\omega^2S(\vec{x})-c_0^2 \Delta S(\vec{x})+ \rho_0^2 u_1 \Delta^2 S(\vec{x})=0.\label{eq:HelmholtzKorteweg}
\end{equation}
We believe that this equation has not been considered previously in the literature.

We first comment on two limiting regimes of the Helmholtz--Korteweg equation: the nearly incompressible regime, with $u_1\rho_0^2\gg c_0^2$, and the nearly ideal regime, with $u_1\rho_0^2\ll c_0^2$.
As previously discussed, \eqref{eq:Korteweg} can be derived within the hyperelasticity framework, in particular, the elastic energy density associated with \eqref{eq:Korteweg} is given by
\begin{equation}
    \label{eq:elasticEnergy}
    W(\rho, \nabla\rho) = \errata{\frac{1}{2}c_0^2 \rho^2} + \frac{1}{2}u_1 \norm{\nabla \rho}^2.
\end{equation}
This energy shows that as $u_1$ increases, a larger penalty is imposed on the gradient of the density field, and in the limit this imposes a constraint of spatially-constant density.
Thus, for nearly incompressible fluids such as water, we expect a large value of $u_1$. However, the speed of sound $c_0$ also increases with incompressibility. Thus, from \eqref{eq:HelmholtzKorteweg}, the limiting regime is determined by the density $\rho_0$.
With a sufficiently large density $\rho_0$ and $\omega^2$ comparable to $\rho_0^2u_1$, the Helmholtz--Korteweg equation \eqref{eq:HelmholtzKorteweg} reduces to the time-harmonic Kirchhoff--Love equation, i.e.
\begin{equation}
    \Delta^2 S(\vec{x})-\mu S(\vec{x}) = 0, \qquad \mu =\frac{\omega^2}{c_0^2u_1\rho_0^2}\;.\label{eq:KirchhoffLove}
\end{equation}

The previous equation suggests that sufficiently dense and incompressible Korteweg fluids will behave as thin fluid shells at sufficiently high frequencies.
On the other hand, for compressible fluids we expect that both $u_1$ and $c_0$ will be smaller.
Hence, for a sufficiently rarefied Korteweg fluid we expect to retrieve the classical Helmholtz equation governing time-harmonic acoustic waves in compressible fluids, i.e.
\begin{equation}
    \Delta S(\vec{x}) + \mu S(\vec{x}) = 0, \qquad \mu=\frac{\omega^2}{c_0^2}\;.
\end{equation}

We must also consider boundary conditions.
Since the Helmholtz--Korteweg equation is a fourth-order partial differential equation, we need to impose two boundary conditions.
Furthermore, for the Helmholtz--Korteweg equation to be consistent with the classical Helmholtz equation in the nearly ideal regime, we would like to consider boundary conditions that incorporate the ones imposed for the classical Helmholtz equation.
We will here consider three types of boundary conditions that might occur when the fluid is in contact with an obstacle along an interface $\Gamma$.
\begin{enumerate}
    \item Sound-soft boundary conditions, which correspond to imposing that the acoustic pressure vanishes along $\Gamma$.
    From \eqref{eq:pressureKorteweg} we know that the acoustic pressure is given by
    \begin{equation}
        c_0^2 \rho_0 S(\vec{x}) - \rho_0^3 u_1 \Delta S(\vec{x}) = 0.
    \end{equation}
    \rev{Thus, assuming the sound-soft boundary conditions for the classical Helmholtz equations also apply, we obtain that the sound-soft boundary conditions correspond to imposing homogeneous Dirichlet boundary conditions on $S(\vec{x})$ and on $\Delta S(\vec{x})$.}
\item Sound-hard boundary conditions, which correspond to imposing that the normal \errata{component} of the fluid velocity $\errata{\vec{\nu}\cdot \vec{v}}$ vanishes along $\Gamma$.
    Using \eqref{eq:linearisedBalanceLaw} and assuming the fluid velocity is also time-harmonic, we get
    \begin{equation}
	    \errata{-i\omega \rho_0 (\vec{V}\cdot \vec{\nu})} =  c_0^2 \partial_{\vec{\nu}} S(\vec{x}) - \rho_0^2 u_1 \partial_{\vec{\nu}} \Delta S(\vec{x}).
    \end{equation}
    \rev{Thus, assuming that sound-hard boundary conditions for the classical Helmholtz equation also apply, we obtain that the sound-hard boundary conditions correspond to imposing homogeneous Neumann boundary conditions on $S(\vec{x})$ and on $\Delta S(\vec{x})$.}
\item Impedance boundary conditions, which correspond to imposing that the normal \errata{component} of the fluid velocity $\errata{\vec{\nu}\cdot \vec{v}}$ is proportional to the excess pressure along $\Gamma$.
	\rev{Imposing the acoustic pressure \eqref{eq:pressureKorteweg} to be proportional to $\errata{\vec{\nu} \cdot \vec{v}}$, computed from \eqref{eq:linearisedBalanceLaw} using the assumption that the fluid velocity is also time-harmonic, we obtain
    \begin{align}
	    c_0^2 \rho_0 S(\vec{x}) - \rho_0^3 u_1 \Delta S(\vec{x}) &= \errata{i\zeta}c_0^2 \partial_{\vec{\nu}} S(\vec{x}) \\
								     &- \errata{i\zeta}\rho_0^2 u_1 \partial_{\vec{\nu}} \Delta S(\vec{x}).\nonumber
    \end{align}
    }
    \rev{Assuming the impedance boundary conditions for the classical Helmholtz equation also apply, we observe that impedance boundary conditions are equivalent to imposing Robin boundary conditions on $S(\vec{x})$ and on $\Delta S(\vec{x})$.}
    The proportionality constant is usually denoted by $\zeta$ and is called the impedance of the boundary.
\end{enumerate}
\rev{Sound soft boundary conditions model obstacles that fully absorb the amplitude of the incoming wave, such as anechoic chambers. Sound hard boundary conditions model obstacles that fully reflect the incoming wave, such as rigid walls. Impedance boundary conditions model obstacles that partially absorb and partially reflect the incoming wave, such as porous materials.}
\section{Nematic Helmholtz--Korteweg equation}

We now consider nematic Korteweg fluids, with stress tensor given by \eqref{eq:NematicKorteweg}.
Virga has studied the propagation of sound waves in the most prominent such fluids, nematic liquid crystals~\cite{VirgaNematoacoustics}.
Historically the interaction of acoustic waves with the nematic director field was first explained by means of the minimal entropy production principle~\cite{DionJacob, Dion}, i.e.~the acoustic anisotropy is assumed to be the result of calamitic molecules reorienting in order to minimize the propagation losses.
We here assume the aligning torque acting on the nematic director field is of elastic nature, rather than of a dissipative viscous one.
This idea was already proposed, and validated experimentally, by Mullen, L\"uthi, and Stephen \cite{MullenLuti}. These authors augment the elastic energy density with a term representing the interaction of the nematic director field with a planar acoustic wave with wave-vector $\vec{k}$, i.e.
\begin{equation}
    W_a(\vec{k},\vec{n}) = c_1\abs{\vec{k}}^2 + c_2\abs{\vec{k}\cdot \vec{n}}^2,
\end{equation}
where $c_1, c_2 > 0$ are constitutive constants.
This approach was extended to the case of a general acoustic wave in Selinger et al.~\cite{SelingerEtAll}, where the elastic energy augmentation is assumed to couple the nematic director field with the gradient of the density field.
In this context Virga's theory regards liquid crystals as Korteweg fluids and proposes the constitutive relation \eqref{eq:NematicKorteweg} for the Cauchy stress tensor, which in a hyperelastic framework can be derived from the elastic energy density \rev{$W(\rho, \nabla\rho, \vec{n})$} \cite{Capriz}.
The acoustic energy $W_a$ is derived from $W$ by assuming that the density $\rho$ represents a planar wave.

Assuming the condensation $s(\vec{x},t)$ is time-harmonic, as in \eqref{eq:timeHarmonic}, we can expand $\nabla\cdot \underline{\underline{\sigma}}$ and drop any term of order $\mathcal{O}(\varepsilon^2)$ to get
\begin{align}
    \nabla\cdot \underline{\underline{\sigma}}\approx \Re\Big[-\rho_0c_0^2\nabla S(\vec{x}) &+ u_1\rho_0^3 \nabla (\Delta S(\vec{x}))\\
    &+\rho_0^3 u_2 \nabla\left((\nabla S\cdot \vec{n})\vec{n}\right)\Big].\nonumber
\end{align}
Substituting this expression in \eqref{eq:timeHarmonicGenWave} and dividing by $\rho_0 e^{-\omega t}$ we get
\begin{align}
    -\omega^2S(\vec{x})-c_0^2\Delta S(\vec{x}) &+ \rho_0^2 u_1 \Delta^2 S(\vec{x})\label{eq:NematicKortewegUnExp}\\
    & + \rho_0^2 u_2 \nabla\cdot \nabla\left[\nabla\cdot ((\nabla S\cdot \vec{n})\vec{n})\right] = 0.\nonumber
\end{align}
We now focus our attention on the last term of the previous equation, which can be expanded, \uz{using \eqref{eq:vectorCalculusIdentity},} as
\begin{align}
    \nabla\cdot \nabla\left[\nabla\cdot ((\nabla S\cdot \vec{n})\vec{n})\right] = \nabla\cdot &\nabla\Big[\underline{\underline{\mathcal{H}S}}\vec{n}\cdot \vec{n} \label{eq:nematicExp}\\
    &+ \underline{\underline{\nabla \vec{n}}}\nabla S\cdot \vec{n} \nonumber \\
    &+ (\nabla S\cdot \vec{n})(\nabla \cdot \vec{n})\Big],\nonumber
\end{align}
where $\underline{\underline{\mathcal{H}S}}$ is the Hessian matrix of $S$.
Substituting this expansion in \eqref{eq:NematicKortewegUnExp} yields,
\begin{align}
    -\omega^2S(\vec{x})-c_0^2\Delta S(\vec{x}) + \rho_0^2 u_1 \Delta^2 &S(\vec{x})\label{eq:NematicKortewegExp}\\
     + \rho_0^2 u_2 \nabla\cdot &\nabla\Big[\underline{\underline{\mathcal{H}S}}\vec{n}\cdot \vec{n} \nonumber \\
    &+ \underline{\underline{\nabla \vec{n}}}\nabla S\cdot \vec{n} \nonumber \\
    &+ (\nabla S\cdot \vec{n})(\nabla \cdot \vec{n})\Big] = 0.\nonumber
\end{align}
A reasonable assumption to proceed, as done in \cite{VirgaNematoacoustics}, is that the nematic director field $\vec{n}$ is regarded as undistorted at the acoustic length scale, so that we can assume $\underline{\underline{\nabla \vec{n}}} = 0$.
Under this hypothesis, \eqref{eq:nematicExp} simplifies to
\begin{align}
    \nabla\cdot \nabla\left[\nabla\cdot ((\nabla S\cdot \vec{n})\vec{n})\right] = \nabla\cdot \nabla\left[\vg{\vec{n}\cdot\underline{\underline{\mathcal{H}S}}\vec{n}}\right].\label{eq:nematicSimplified}
\end{align}
This yields the nematic Helmholtz--Korteweg equation, i.e.
\begin{align}
    -\omega^2S(\vec{x})-c_0^2\Delta S(\vec{x}) + &\rho_0^2 u_1 \Delta^2 S(\vec{x})\label{eq:NematicHelmholtzKorteweg}\\
    + &\rho_0^2 u_2 \nabla\cdot \nabla\left[\vg{\vec{n}\cdot\underline{\underline{\mathcal{H}S}}\vec{n}}\right] = 0.\nonumber
\end{align}
\rev{Under this hypothesis, and the additional assumption that the degree of orientational order is constant, describing the nematic order with a director field $\vec{n}$ or a $Q$-tensor are equivalent.}

We would like to comment on a secondary effect of the assumption that the nematic director field $\vec{n}$ is undistorted at the acoustic length scale.
In particular, the assumption $\underline{\underline{\nabla \vec{n}}} = 0$ implies that any acoustic effect associated to the Oseen--Frank portion of the elastic energy density, i.e.
\begin{equation}
    W_{OF}(\underline{\underline{\nabla \vec{n}}}) = \frac{1}{2}K \left(\underline{\underline{\nabla \vec{n}}}\,:\underline{\underline{\nabla \vec{n}}}\right),
\end{equation}
is neglected. (Here we have adopted the one-constant approximation for the Oseen--Frank elastic energy density, where $K > 0$ is the Frank elastic constant.)
This should come as no surprise, since the Oseen--Frank elastic energy density and the corresponding Leslie--Ericksen equations are usually derived under the assumption that the fluid under consideration is incompressible.
In \cite{FarrellEtAll} an inviscid compressible variant of the Leslie--Ericksen equations has been derived from kinetic considerations.
In particular, the Euler equations are augmented with a compressible Ericksen tensor, i.e.
\begin{equation}
    \label{eq:EricksenAugmentation}
    \underline{\underline{\sigma}}^{(E)}\coloneqq \lambda p \, \underline{\underline{\nabla \vec{n}}}^T\underline{\underline{\nabla\vec{n}}},
\end{equation}
where $\lambda > 0$ is a constitutive parameter uniquely determined by the shape of the calamitic molecules constituting the liquid crystal.
Over a longer length scale than the acoustic one we will need also to take into account the effect of the augmentation term \eqref{eq:EricksenAugmentation}.

Lastly, we need to discuss boundary conditions for the nematic Helmholtz--Korteweg equation.
The boundary conditions for the nematic case differ slightly from the isotropic Helmholtz--Korteweg equation:
\begin{enumerate}
    \item Sound-soft boundary conditions will change since excess pressure is now defined, from \eqref{eq:pressureNematic}, as
    \begin{equation}
        c_0^2 \rho_0 S(\vec{x}) - \rho_0^3 u_1 \Delta S(\vec{x}) - u_2\rho_0^3 \left(\vec{n}\cdot \underline{\underline{\mathcal{H}S}}\vec{n}\right)= 0.
    \end{equation}
    Sound-soft boundary conditions thus correspond to imposing homogeneous Dirichlet boundary conditions on $S(\vec{x})$ and
    \begin{equation}
        \Delta S(\vec{x}) = -\frac{u_2}{u_1}\left( \vec{n}\cdot \underline{\underline{\mathcal{H}S}}\vec{n}\right).
    \end{equation}
	\item Sound-hard boundary conditions also change since the normal \errata{component} of the fluid velocity $\errata{\vec{\nu}\cdot\vec{v}}$ now satisfies the equation
    \begin{align}
	    \errata{-i\omega \rho_0 (\vec{v}\cdot \vec{\nu})} =  c_0^2 \partial_{\vec{\nu}} S(\vec{x}) &- \rho_0^2 u_1 \partial_{\vec{\nu}} \Delta S(\vec{x}) \\
        &- \rho_0^2 u_2 \partial_{\vec{\nu}}\left(\vec{n}\cdot \underline{\underline{\mathcal{H}S}}\vec{n}\right).
    \end{align}
    Sound-hard boundary conditions thus correspond to imposing homogeneous Neumann boundary conditions on $S(\vec{x})$ and
    \begin{equation}
        \partial_{\vec{\nu}}\,\Delta S(\vec{x}) = -\frac{u_2}{u_1}\partial_{\vec{\nu}}\left( \vec{n}\cdot \underline{\underline{\mathcal{H}S}}\vec{n}\right).
    \end{equation}
    \item Some computation shows that the impedance boundary conditions for the nematic Helmholtz--Korteweg equation are equivalent to imposing Robin boundary conditions on $S(\vec{x})$ and
    \begin{align}
        \partial_{\vec{\nu}} \, \Delta S(\vec{x}) = i\zeta \Delta S(\vec{x}) &+ i\zeta \frac{u_2}{u_1}\left( \vec{n}\cdot \underline{\underline{\mathcal{H}S}}\vec{n}\right) \nonumber \\
        &- \frac{u_2}{u_1}\partial_{\vec{\nu}}\left( \vec{n}\cdot \underline{\underline{\mathcal{H}S}}\vec{n}\right).
    \end{align}
\end{enumerate}

\section{Plane waves}
\label{sec:planeWaves}
We wish to build intuition about the Helmholtz--Korteweg and nematic Helmholtz--Korteweg equations. We also wish to verify their physical correctness, by checking that the dispersion relation they imply matches that derived by Virga~\cite{VirgaNematoacoustics,VirgaSonnet} using the continuity equation and the balance law of linear momentum.
We therefore consider plane wave solutions, given by
\begin{equation}
    \label{eq:planeWaves}
    S(\vec{x}) = s_0e^{ik(\vec{x}\cdot\vec{d})},
\end{equation}
where $s_0=\mathcal{O}(\varepsilon)$, $\vec{d}$ is the unit vector that prescribes the direction in which the wave propagates, $k$ is the wave-number and the wave-vector $\vec{k}$ is given by $\vec{k} \coloneqq k\vec{d}$.
Substituting \eqref{eq:planeWaves} in \eqref{eq:HelmholtzKorteweg} and dividing by $e^{i\vec{k}\cdot\vec{x}}$ we obtain the dispersion relation
\begin{equation}
    \label{eq:quadraticEquation}
    -\omega^2+c_0^2k^2 + \rho_0^2 u_1 k^4 = 0.
\end{equation}

We now introduce the nondimensional wave-vector defined as
\begin{equation}
    \label{eq:nondimensionalWaveVector}
    \vec{\kappa} \coloneqq \frac{c_0}{\omega}\vec{k}.
\end{equation}
Using the nondimensional wave-vector and the corresponding wave-number $\kappa$ we can rewrite \eqref{eq:quadraticEquation} as
\begin{equation}
    \label{eq:nondimensionalQuadraticEquation}
    -1+\kappa^2 + \frac{1}{4}\tau_1^2\omega^2\kappa^4= 0,\qquad
    \tau_1\coloneqq 4\frac{\rho_0\sqrt{u_1}}{c_0^2}.
\end{equation}
Here $\tau_1$ represents the characteristic time scale of Korteweg acoustic waves, i.e.~the time scale over which the Korteweg nature of the fluid becomes relevant in the propagation of acoustic waves.
Clearly both \eqref{eq:quadraticEquation} and \eqref{eq:nondimensionalQuadraticEquation} have solutions with both real and imaginary parts. In particular \eqref{eq:nondimensionalQuadraticEquation} has solutions
\begin{equation}
    \label{eq:nondimensionalWaveNumbers}
    \kappa = \pm \frac{1}{\tau_1\omega}  \left[-2 \pm 2\sqrt{1+\tau_1^2\omega^2}\right]^{\frac{1}{2}}.
\end{equation}
Since the wave-numbers $\kappa$ can have both real and imaginary parts, we can have both propagating and evanescent waves, i.e.~waves that oscillate in the direction of the real part of $\vec{\kappa}$ and waves that decay exponentially in the direction of the imaginary part of $\vec{\kappa}$.

Following \cite{VirgaNematoacoustics,VirgaSonnet}, by analogy to the Helmholtz equation we assume that the wave-number $k$ can be decomposed as
\begin{equation}
    \label{eq:decomposition}
    k = k^{(R)} + ik^{(I)}, \qquad k^{(R)}=\frac{\omega}{c},
\end{equation}
{where $k^{(R)}, k^{(I)} \in \mathbb{R}$ and where $c>0$ is the effective speed of sound, i.e.~the actual (and possibly anisotropic) speed of sound with which the wave propagates, as opposed to the isotropic speed of sound $c_0$.}
We then introduce a new dimensionless quantity $\kappa^{(I)}$, i.e.
\begin{equation}
    \label{eq:nondimensional2}
    \kappa^{(I)}\coloneqq \frac{c_0}{\omega}k^{(I)}.
\end{equation}
Using \eqref{eq:nondimensional2} we can split \eqref{eq:nondimensionalQuadraticEquation} in real and imaginary parts. Considering first the imaginary part, we find two non-trivial solutions for $\kappa^{(I)}_0$, i.e.
\begin{equation}
    \label{eq:nondimensionalImaginary}
    \kappa^{(I)} = \pm \sqrt{\frac{2}{\tau_1^2\omega^2}-\left(\frac{c_0}{c}\right)^2}.
\end{equation}
Substituting \eqref{eq:nondimensionalImaginary} in the real part of \eqref{eq:nondimensionalQuadraticEquation}, {we focus our attention on solutions that are} either purely real or purely imaginary.

We first consider the purely real case, where $\kappa^{(I)}_0$ vanishes, and $k=\omega/c$.
We express \eqref{eq:quadraticEquation} using the characteristic time scale $\tau_1$ to get
\begin{equation}
    \label{eq:quadraticEquation2}
    -\omega^2+c_0^2k^2 + \frac{1}{4}\tau_1^2c_0^4k^4= 0.
\end{equation}
Substituting our expression for $k$ and dividing by $\omega^2$ yields
\begin{equation}
    \label{eq:quadraticEquation3}
    -1+\left(\frac{c_0}{c}\right)^2+ \frac{1}{4}\tau_1^2\omega^2\left(\frac{c_0}{c}\right)^4= 0.
\end{equation}
Solving this equation for $c_0/c$ we find only one real positive root, i.e.
\begin{equation}
    \label{eq:speedOfSound}
    \frac{c_0}{c} = \omega\tau_1\left[2\left(\sqrt{1+\omega^2\tau_1^2}-1\right)\right]^{-\frac{1}{2}}.
\end{equation}
Thus, as $\omega\tau_1$ increases, the speed of a propagating acoustic wave in a Korteweg fluid increases as well.
This matches~\cite[eq.~(82)]{VirgaNematoacoustics}.

We next consider the other case where $k$ is purely imaginary, i.e.~$k = ik^{(I)}$.
Inspired by the phenomenon of total internal reflection, we will assume
\begin{equation}
    \label{eq:totalInternalReflection}
    k^{(I)} = - \frac{\omega}{c}\sqrt{\alpha}, \qquad \alpha \geq 0.
\end{equation}
{
The real number $\alpha$ has a physical interpretation in the context of total internal reflection, related to the angle of incidence, and the refractive indices of the two media involved.
}
Substituting \eqref{eq:totalInternalReflection} in \eqref{eq:quadraticEquation} we find
\begin{equation}
    \label{eq:totalInternalReflectionDispersion}
    -1-\left(\frac{c_0}{c}\right)^2\alpha+ \frac{1}{4}\tau_1^2\omega^2\left(\frac{c_0}{c}\right)^4\alpha^2= 0.
\end{equation}
Solving for $\alpha$ we find two solutions, i.e.
\begin{equation}
    \label{eq:totalInternalReflectionRoots}
    \alpha = 4\left(\frac{c}{c_0}\right)^2 \left[\frac{-1\pm \sqrt{1+\omega^2\tau_1^2}}{\omega^2\tau_1^2}\right],
\end{equation}
The penetration depth $\delta$ of the evanescent wave, i.e.~the distance over which the amplitude of the wave decays by a factor of $1/e$, is the absolute value of the reciprocal of the imaginary part of the wave-number.
We begin substituting \eqref{eq:totalInternalReflectionRoots} in \eqref{eq:totalInternalReflection} and discarding the imaginary part of $\sqrt{\alpha}$ to get 
\begin{equation}
    \label{eq:penetrationDepth}
    \delta = \frac{c_0}{2\omega}\left[\frac{-1+\sqrt{1+\tau_1^2\omega^2}}{\tau_1^2\omega^2}\right]^{-\frac{1}{2}}
\end{equation}
Asymptotically for $\omega\tau_1\gg 1$ we can approximate \eqref{eq:penetrationDepth} as
\begin{equation}
    \label{eq:penetrationDepthApprox}
    \delta \approx \frac{c_0}{2\omega}\sqrt{\tau_1\omega},
\end{equation}
which suggests that the penetration depth of an evanescent wave is inversely proportional to the square root of $\omega$ and directly proportional to the square root of $\tau_1$.

\begin{figure}[h]
    \centering
    \includegraphics[width=0.235\textwidth]{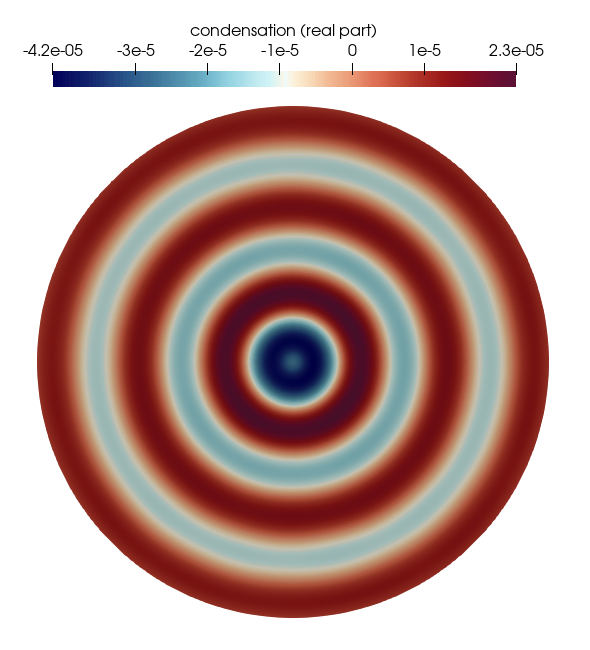}
    \includegraphics[width=0.235\textwidth]{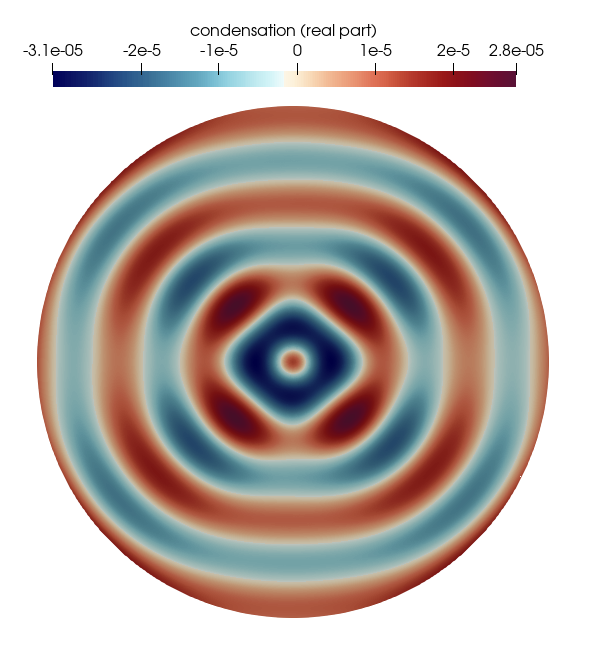}
    $\newline$
    $\newline$
    \includegraphics[width=0.235\textwidth]{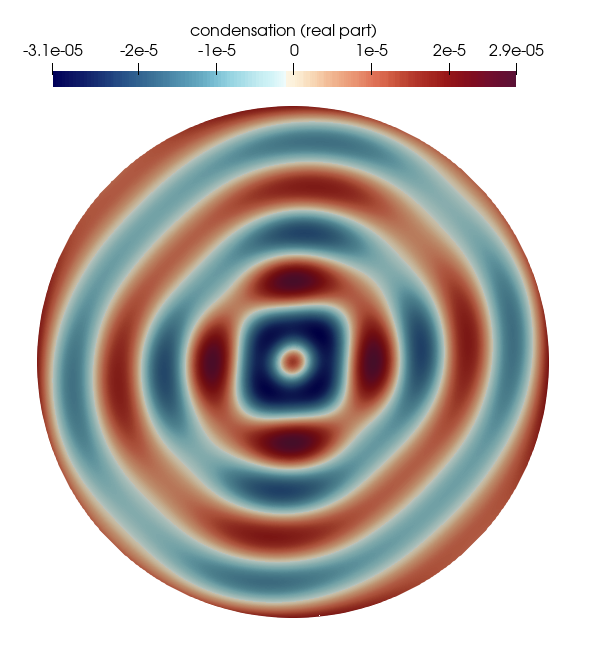}
    \includegraphics[width=0.235\textwidth]{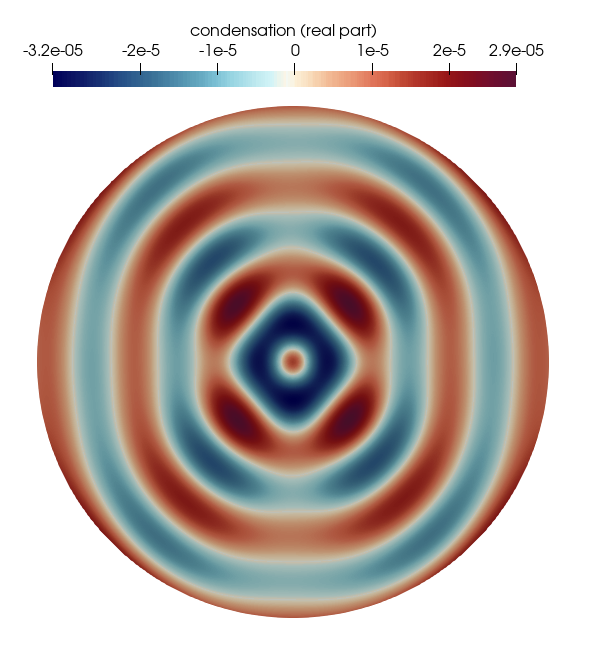}
    \caption{The condensation wave satisfying the Helmholtz--Korteweg equation (top left) and the nematic Helmholtz--Korteweg equation with different prescribed nematic directors, i.e.~parallel to the $x$-axis (top right), parallel to the diagonal (bottom left) and parallel to the $y$-axis (bottom right).
    Each simulation uses as initial condition a Gaussian pulse centered in the middle of the domain.}
    \label{fig:speedOfSound}
\end{figure}

Lastly we would like to consider the nematic Helmholtz--Korteweg equation \eqref{eq:NematicHelmholtzKorteweg} and the corresponding plane wave solutions.
Substituting \eqref{eq:planeWaves} in \eqref{eq:NematicHelmholtzKorteweg} we obtain
\begin{equation}
    \label{eq:NematicQuadraticEquation}
    -\omega^2+c_0^2k^2 + \rho_0^2 u_1 k^4 + \rho_0^2 u_2 k^4\left(\vec{d}\cdot \vec{n}\right)^2= 0,
\end{equation}
where we have used the fact that $\underline{\underline{\mathcal{H}S}} = k^2 \underline{\underline{S(\vec{d}\otimes\vec{d})}}$.
This matches the dispersion relation derived by Virga~\cite[eq.~(78)]{VirgaNematoacoustics}, in the inviscid regime.
Collecting the fourth order terms in \eqref{eq:NematicQuadraticEquation} and defining $\xi$ as the angle between $\vec{d}$ and $\vec{n}$ we find
\begin{equation}
    \label{eq:NematicQuadraticEquation2}
    -\omega^2+c_0^2k^2 + \left[\rho_0^2 u_1  + \rho_0^2 u_2 \cos^2(\xi)\right]k^4= 0.
\end{equation}
We now introduce the nematic Korteweg characteristic time scale $\tau_2$, defined as 
\begin{equation}
    \label{eq:tau2}
    \tau_2 \coloneqq 4\frac{\rho_0\sqrt{u_1+u_2\cos^2(\xi)}}{c_0^2}.
\end{equation}
Notice that due to the presence of the $\cos^2(\xi)$ term in the definition of $\tau_2$, the characteristic time scale $\tau_2$ is anisotropic.
Furthermore we can rewrite \eqref{eq:NematicQuadraticEquation2} as
\begin{equation}
    \label{eq:NematicQuadraticEquation3}
    -1+\kappa^2 + \frac{1}{4}\tau_2^2\omega^2\kappa^4= 0,
\end{equation}
with roots
\begin{equation} \label{eq:nondimensionalWaveNumbersNematic}
    \kappa = \pm \frac{1}{\tau_2\omega}  \left[-2 \pm 2\sqrt{1+\tau_2^2\omega^2}\right]^{\frac{1}{2}}.
\end{equation}
Proceeding in a manner analogous to the one used for the Helmholtz--Korteweg equation we find that for propagating waves the dispersion relation is given by
\begin{equation}
    \label{eq:NematicSpeedOfSound}
    \frac{c_0}{c} = \omega\tau_2\left[2\left(\sqrt{1+\omega^2\tau_2^2}-1\right)\right]^{-\frac{1}{2}},
\end{equation}
which due to the anisotropic nature of $\tau_2$ is also anisotropic. In particular, the speed of sound of a propagating acoustic wave in a nematic Korteweg fluid is greatest when the wave propagates along the nematic director field, as observed in the experiment of Mullen, L\"uthi \& Stephen \cite{MullenLuti}.
The anisotropic nature of the speed of sound in nematic Korteweg fluids is observed in our numerical simulations, as can be seen in Figure \ref{fig:speedOfSound} \footnote{Since the Helmholtz--Korteweg equation is fourth order, we employ $C^1$-conforming Argyris finite elements~\cite{Argyris} in Firedrake~\cite{firedrake}. The PDE analysis for these equations, and a numerical analysis of their discretisations, will be presented elsewhere.}.

Furthermore, we conclude that the penetration depth of an evanescent wave in a nematic Korteweg fluid is greatest when the wave propagates along the nematic director field.
We conjecture that this effect could be observed experimentally in a modification of the Mullen--L\"{u}thi--Stephen experiment.
\section{Plane wave reflection}
\label{eq:planeWaveReflection}
We consider $\mathbb{R}^2$ where the upper half-plane is occupied by a Korteweg fluid, and the real axis is the interface between the fluid and an obstacle.
Let $S^-$ denote an incoming plane wave of the form \eqref{eq:planeWaves}, i.e.
\begin{equation}
    \label{eq:incomingWave}
    S^-(\vec{x}) = s_0e^{ik\left(d_1x_1 + d_2x_2\right)},\qquad -d_1,d_2\leq 0.
\end{equation}
Since the real axis is the interface between the fluid and the obstacle, by Snell's reflection law we have a reflected wave of the form
\begin{equation}
    \label{eq:reflectedWave}
    S^+(\vec{x}) = s_0Ae^{ik\left(d_1x_1 - d_2x_2\right)},
\end{equation}
where $A$ is the amplitude.
Since $S^+$ and $S^-$ are solutions of \eqref{eq:HelmholtzKorteweg} or \eqref{eq:NematicHelmholtzKorteweg}, by the linearity of these equations $S\coloneqq S^+ + S^-$ is also a solution.
Furthermore, it is an easy exercise to show that the boundary conditions prescribed on $S$ impose a constraint on the amplitude $A$ of the reflected wave, i.e.
\begin{enumerate}
    \item For sound-soft boundary conditions along the real axis we have
    $s_0e^{ikd_1x_1} + s_0Ae^{ikd_1x_1} = 0$, thus we need to impose $A = -1$.
    \item For sound-hard boundary conditions along the real axis we have $s_0 ik d_2 e^{ikd_1x_1}- s_0 ik d_2 Ae^{ikd_1x_1} = 0$, thus we need to impose $A = 1$.
    \item For impedance boundary conditions along the real axis we have 
    \begin{equation}
        iA(d_2k-\zeta) e^{ikd_1x_1} = -i(d_2k+\zeta) e^{ikd_1x_1},
    \end{equation}
    thus we need to impose $A = k\frac{d_2+\zeta k^{-1}}{d_2-\zeta k^{-1}}$.
\end{enumerate}

First we focus our attention on the case where $k \in \mathbb{R}$.
Using our previous results, we know that the wave-number $k$ of the Helmholtz--Korteweg equation is constant if the Korteweg fluid is homogeneous.
Therefore, we can conclude that the qualitative behaviour of the reflection of a plane wave in a Korteweg fluid is the same as in a simple fluid.
Quantitatively, while sound-soft and sound-hard boundary conditions impose the same constraint on the amplitude of the reflected wave in both Korteweg and simple fluids, impedance boundary conditions impose a different constraint on the amplitude of the reflected wave in Korteweg fluids since the wave-number $k$ also depends on $u_1$, via $\tau_1$.

For the nematic Helmholtz--Korteweg equation the situation is more interesting. While sound-soft and sound-hard boundary conditions impose the same constraint on the amplitude of the reflected wave in both Korteweg and nematic Korteweg fluids, impedance boundary conditions have a qualitatively different effect in nematic Korteweg fluids.
From \eqref{eq:nondimensionalWaveNumbersNematic} we know that the wave-number $k$ is \uz{smaller} when the wave propagates orthogonal to the nematic director field.
Moreover, the amplitude of the reflected wave is \uz{smaller} when the wave propagates orthogonal to the nematic director field.
Thus we can expect that the absorption of sound caused by the presence of impedance boundary conditions is \uz{smaller} if the wave-vector is parallel to the nematic director field.

\begin{figure}
    \centering
    \includegraphics[scale=0.058]{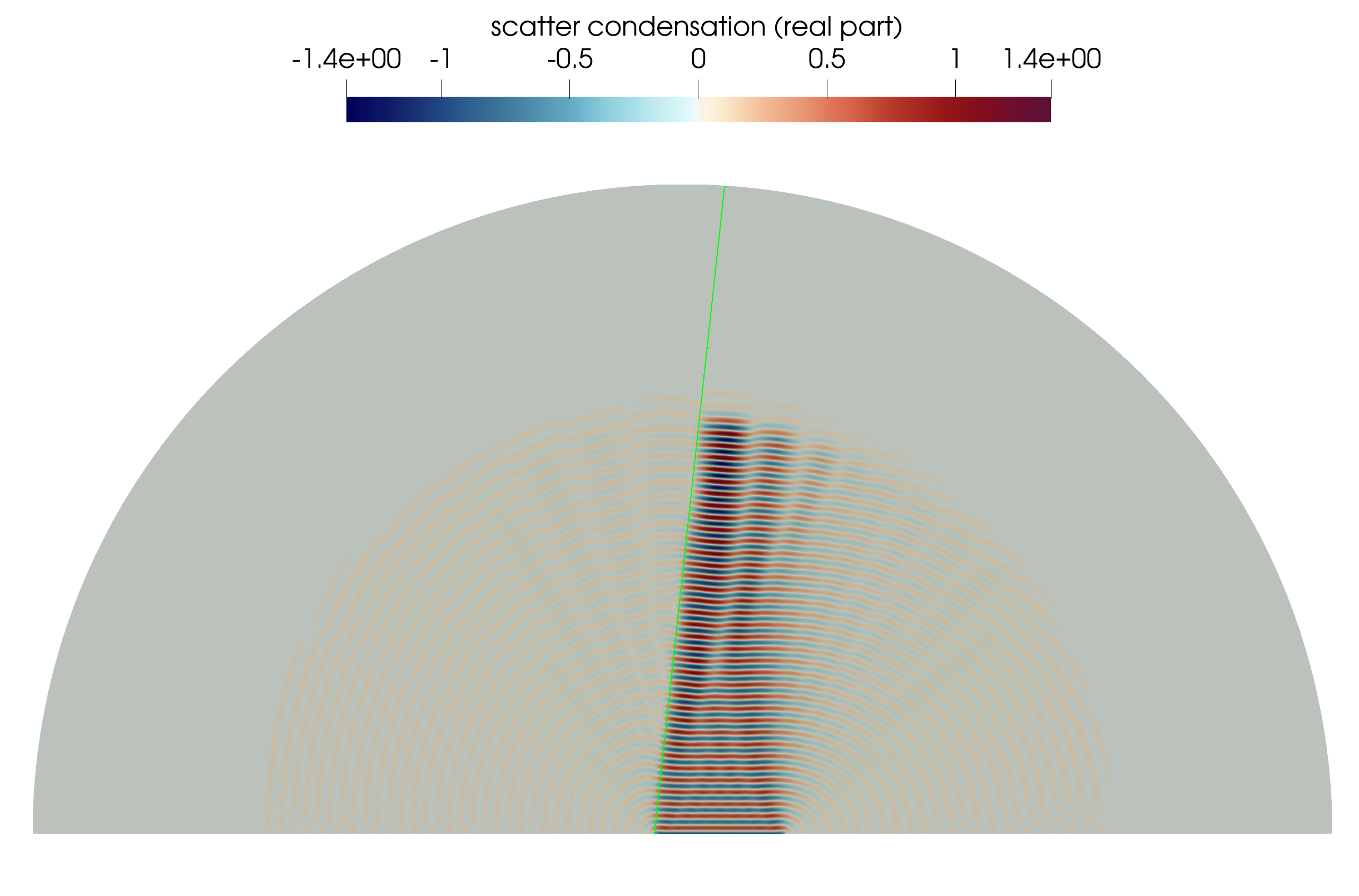}
    \qquad
    \includegraphics[scale=0.058]{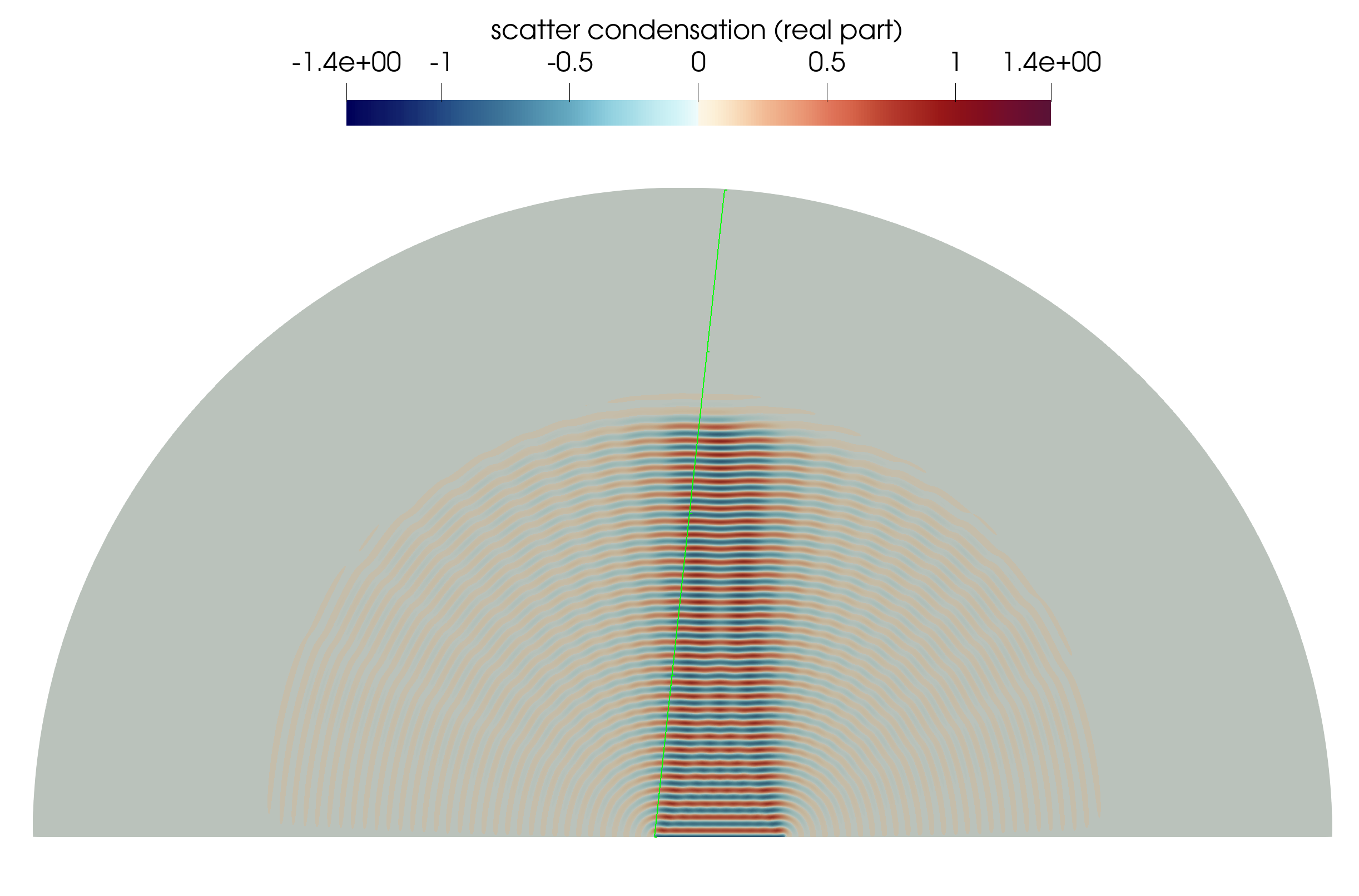}
    \qquad
    \includegraphics[scale=0.069]{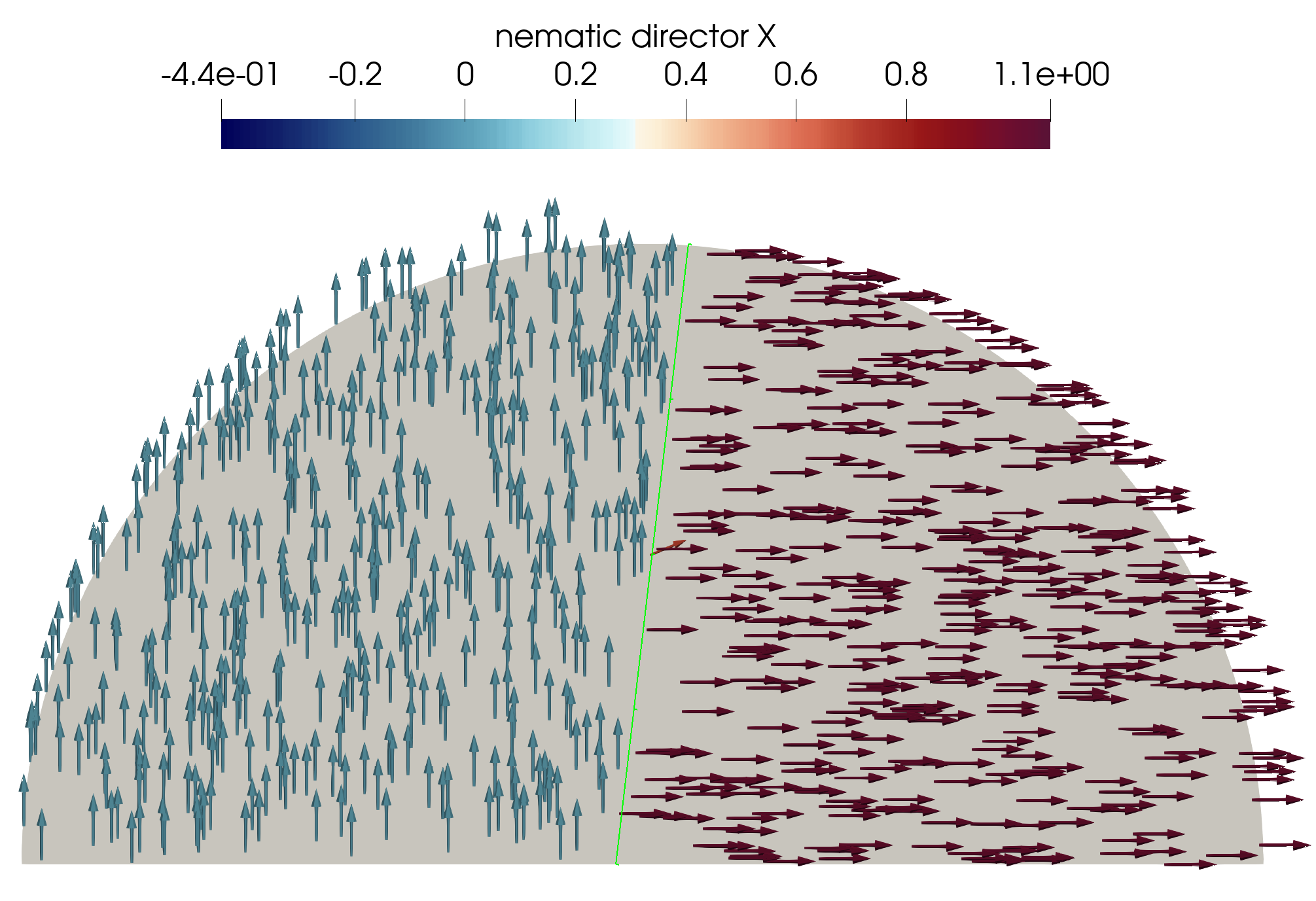}
    \qquad\,
    \includegraphics[scale=0.069]{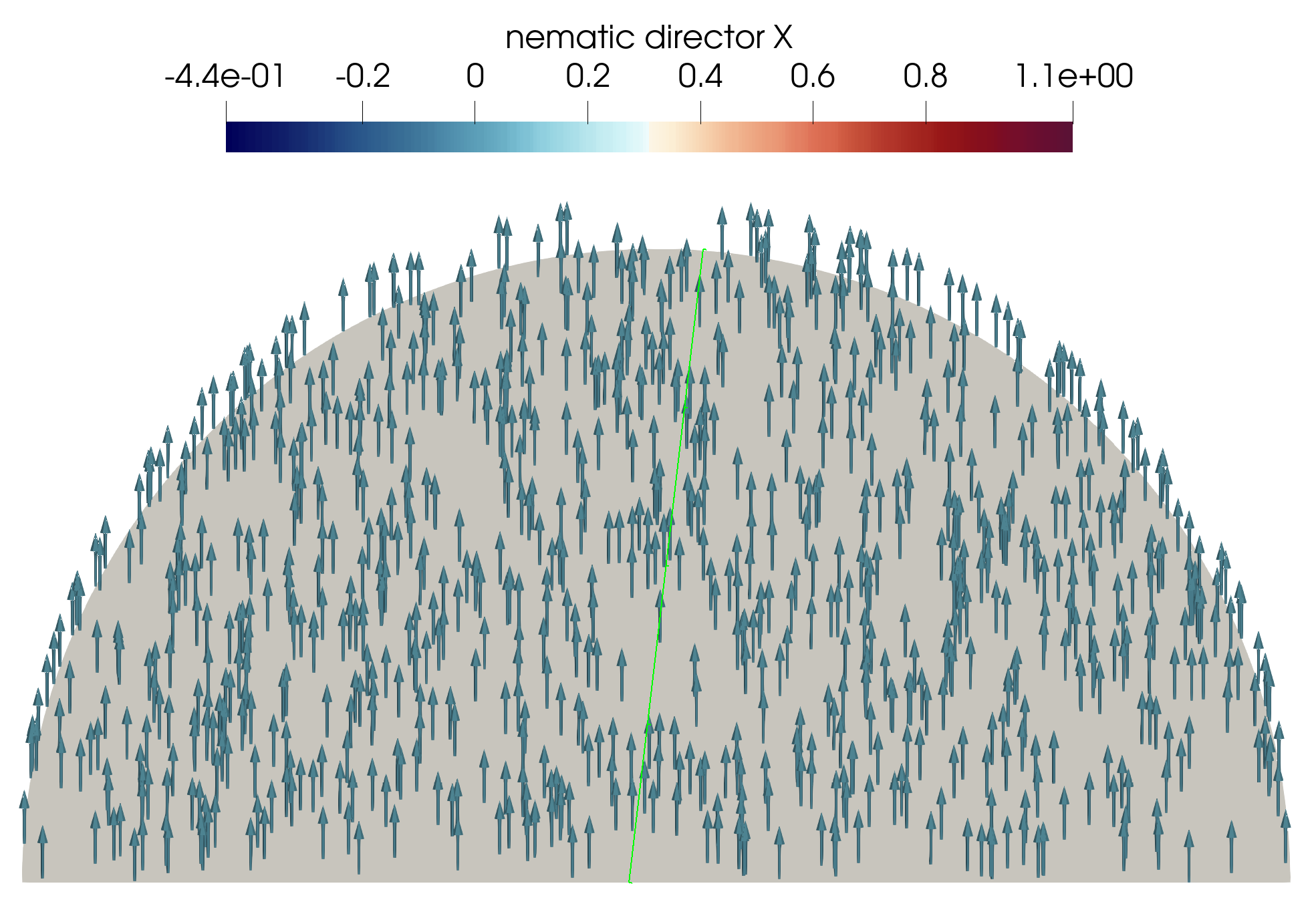}
    \caption{\vg{An acoustic reflection phenomenon in a nematic Korteweg fluid can be caused by a discontinuity in the nematic director field.
    \rev{We consider a Gaussian beam travelling upwards in a semicircular domain, with two different nematic director fields.}
    (Left) The nematic director field is parallel to the wave-vector on the left side of the line marked in green, and orthogonal on the right side. As a result, a plane wave is partially reflected along the line. (Right) No reflection occurs when the nematic director has no discontinuity.}
    \rev{The discontinuity in the nematic director field presented in the left figures cannot be achieved if the nematic director field is a general solution of the Euler--Lagrange equations associated with the Oseen--Frank energy density \cite{Virga}, and should be
    regarded as an approximation of a solution of the Oseen--Frank model exhibiting a sharp transition between two constant orientations.}
    }
    \label{fig:TIR}
\end{figure}

So far we have considered only the case where the wave-number $k$ is real. We now turn our attention to the total internal reflection case, described by imaginary wave-numbers.
Assuming now that the lower half-plane is occupied by another Korteweg fluid, as a consequence of the propagation of $S^+$ we will have a transmitted wave $S^T$ in the lower half-plane.
The transmitted wave will have wave number $k^T = \frac{n_T}{n}k$, where $n_T$ is the refractive index of the lower half-plane and $n$ is the refractive index of the upper half-plane.
Using Snell's law we can compute the direction $\vec{d}^T$ of the transmitted wave, i.e.
\begin{equation}
    \label{eq:transmittedWave}
    d^T_1 = \sin(\theta)\frac{n}{n_T},\qquad d^T_2 = \sqrt{1-\sin^2(\theta)\frac{n^2}{n_T^2}},
\end{equation}
where $\theta$ is the angle of incidence of $S^+$.
When the angle of incidence is greater than the critical angle, i.e.
\begin{equation} 
\theta > \theta_c, \qquad \theta_c = \sin^{-1}\left(\frac{n_T}{n}\right),
\end{equation}
the transmitted wave will undergo total internal reflection, i.e.~$d_2^T$ will be imaginary and the transmitted wave can be expressed as
\begin{equation}
    \label{eq:totalInternalReflectionTransmittedWave}
    S^T(\vec{x}) = s_0e^{i\vec{k}^{T}\cdot\vec{x}}, \qquad \vec{k}^T = \pm
    \begin{bmatrix}
        \sin(\theta)\\
        i\sqrt{\alpha}
    \end{bmatrix}k,
\end{equation}
where $\alpha$ is the real number defined as
\begin{equation}
    \label{eq:alpha}
    \frac{\alpha}{k^2} \coloneqq \left(\frac{n_T}{n}\right)^2-\sin^2(\theta).
\end{equation}
With nematic fluids, since $n$ and $n_T$ depend on the nematic orientation, these observations suggest it may be possible to control whether reflection occurs with electromagnetic fields, as demonstrated in Figure \ref{fig:TIR}.

\section{Scattering by a circular obstacle}
\begin{figure}
    \centering
    \scalebox{0.2}{
        \begin{tikzpicture}
            \node[anchor=north,inner sep=0] at (0,6) {\includegraphics[width=\textwidth]{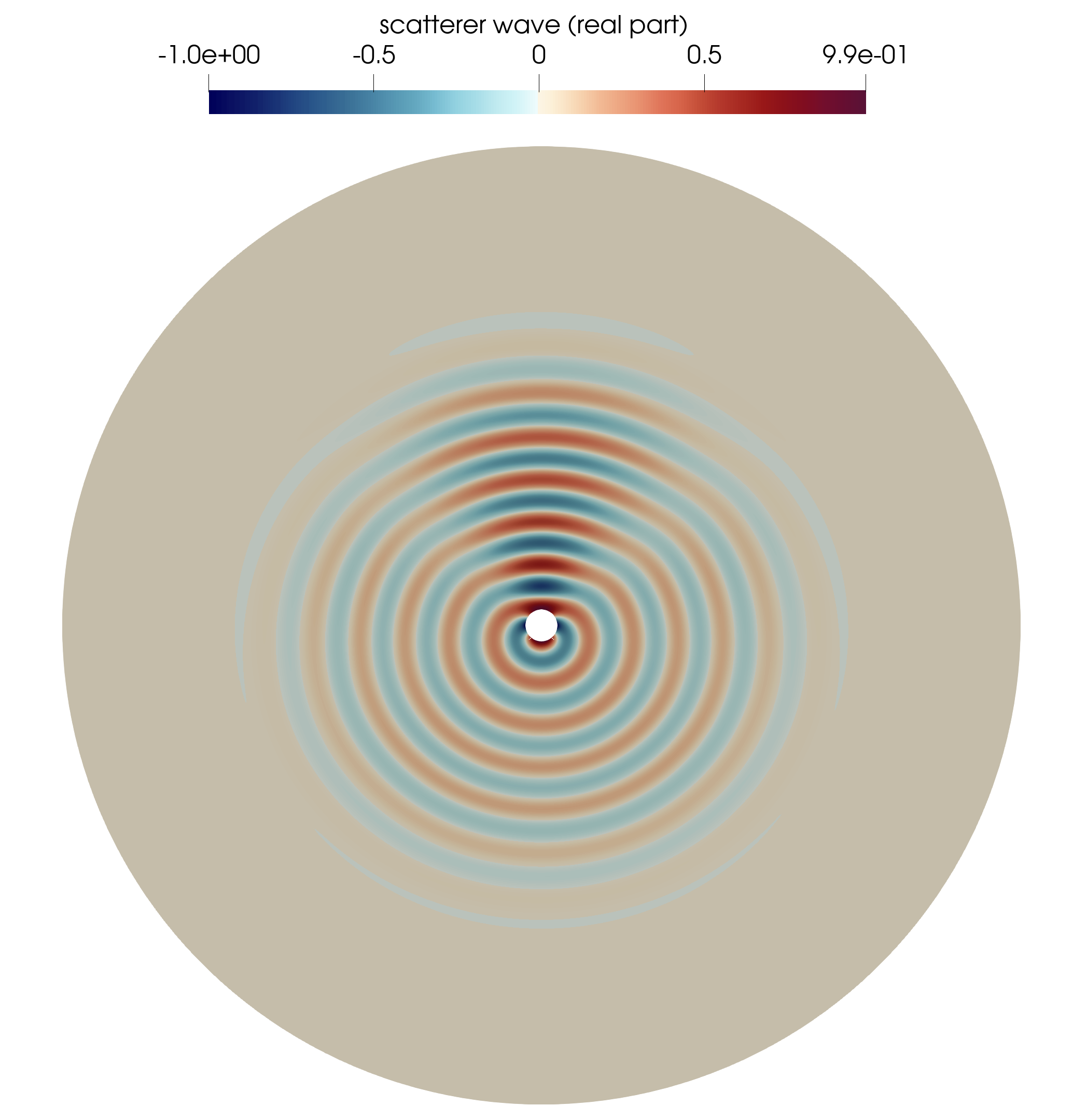}};
            \node [-] (0) at (0, 3) {};
            \node [-] (1) at (0, 0) {};
            \node [-] (2) at (-1, 2) {};
            \node [-] (3) at (1, 2) {};
            \node [-] (4) at (1, 1.75) {};
            \node [-] (5) at (-1, 1.75) {};
            \node [-] (6) at (-1, 1.5) {};
            \node [-] (7) at (1, 1.5) {};
            \node [-] (8) at (-1, 1.25) {};
            \node [-] (9) at (1, 1.25) {};
            \node [-] (10) at (-1, 1) {};
            \node [-] (11) at (1, 1) {};
            \draw [->, ultra thick] (0.center) to (1.center);
            \draw (2.center) to (3.center);
            \draw (5.center) to (4.center);
            \draw (6.center) to (7.center);
            \draw (8.center) to (9.center);
            \draw (10.center) to (11.center);
        \end{tikzpicture}
    }
    \begin{tikzpicture}[scale=0.5]
        \begin{axis}
            \addplot[line width=1.1pt,mark=None, color=seabornred] table [x=y, y=F, col sep=comma] {data/pml_0.001_0.0005_n1_0.0_n2_0.0.csv};
            \addplot[line width=1.1pt,mark=None, color=seabornblue] table [x=y, y=F, col sep=comma] {data/pml_0.001_0.0005_n1_1.57_n2_1.57.csv};
            \legend{$\xi=0$,$\xi=\frac{\pi}{2}$}
        \end{axis}
    \end{tikzpicture}
    \caption{The scattered wave produced by a circular obstacle in a nematic Korteweg fluid with $u_1 = 10^{-3}$ and $u_2 = 5\cdot 10^{-4}$, has a \uz{greater} amplitude when the incoming plane wave is orthogonal to the nematic director field. Recall that $\xi$ is the angle between $\vec{d}$ and $\vec{n}$.
    \vg{We simulated a plane wave propagating parallel to the $y$-axis and impinging on a circular obstacle, centered at the origin (left).
    The amplitude of the scattered wave, for different values of $\xi$, is measured along the $y$-axis (right).}
    An adiabatic layer has been used to implement the Sommerfeld radiation condition on the outer boundary~\cite{oskooi2008}.}
    \label{fig:scattering}
\end{figure}

We next consider the scattering of a plane wave by a circular sound soft obstacle, immersed in a nematic Korteweg fluid.
In particular, we will focus our attention on the physical regime associated with nematic liquid crystals where $u_2 \ll u_1$, i.e.
\begin{equation}
    \label{eq:physicalRegimeScattering}
    \rho_0^2 u_1 \approx \ell^2, \; \rho_0^2 u_2 \approx \gamma^{-1}\ell^2, \;\;\text{ with } \gamma \gg 1 \gg \ell.
\end{equation}
We can express the sound soft scattering problem as a boundary value problem for the nematic Helmholtz--Korteweg equation \eqref{eq:NematicHelmholtzKorteweg}, i.e.
\begin{alignat}{3}
    \label{eq:scatteringProblem}
    -\omega^2S^+(\vec{x})&-c_0^2\Delta S^+(\vec{x}) + \ell^2 \Delta^2 S^+(\vec{x}) \nonumber\\
    &+ \gamma^{-1}\ell^2 \nabla\!\cdot\! \nabla\left[\vg{\vec{n}\cdot \underline{\underline{\mathcal{H}S^+}}\vec{n}}\right] \! =\! 0\; &&\abs{\vec{x}}>1,\;\;\,\nonumber\\
    S^+(\vec{x}) &= S^-(\vec{x})\qquad &&\abs{\vec{x}}=1,\;\;\,\nonumber\\
    \Delta (S^+ - S^-) &+ \gamma^{-1}\vec{n}\cdot \underline{\underline{\mathcal{H}}}(S^+ - S^-)\vec{n} = 0 &&\abs{\vec{x}}=1,\;\;\,\nonumber\\
    \abs{\partial_{\abs{\vec{x}}}S^+(\vec{x})&-ikS^+(\vec{x})}=\mathcal{O}(\abs{\vec{x}}^{-\frac{1}{2}}) &&\abs{\vec{x}} \to \infty,
\end{alignat}
where $S^+$ is the scattered wave and $S^-$ is the incoming plane wave. The second and third equations in \eqref{eq:scatteringProblem} are the sound-soft boundary conditions, while the last equation is the Sommerfeld radiation condition, and $k$ is one of the real roots of \eqref{eq:NematicQuadraticEquation}.
We can consider \eqref{eq:scatteringProblem} as a perturbation of the scattering problem for the Helmholtz equation.
It is important to notice that the problem is a singularly perturbed one, since the Helmholtz--Korteweg equation is a fourth order partial differential equation which reduces to a second order partial differential equation in the limit $\ell\to 0$, for which only one set of boundary conditions can be imposed.
For this reason we expect a boundary layer to form around the obstacle. To study the boundary layer we introduce the change of variables $\vec{\xi} = \ell\,\vec{x}$, and rewrite the first equation of \eqref{eq:scatteringProblem} as
\begin{align}
    \label{eq:boundaryLayerExpansion}
    \nonumber
    -\omega^2S^+(\vec{\xi})-\frac{c_0^2}{\ell^2}\Delta S^+(\vec{\xi}) &+ \frac{1}{\ell^2} \Delta^2 S^+(\vec{\xi}) \\
    &+ \frac{\gamma^{-1}}{\ell^2} \nabla\!\cdot\! \nabla\left[\underline{\underline{\mathcal{H}S^+}}\vec{n}\cdot \vec{n}\right] \! =\! 0.
\end{align}
As $\ell\to 0$ we can consider only the dominant terms in \eqref{eq:boundaryLayerExpansion} to derive a partial differential equation for the boundary layer $S_\ell$, i.e.
\begin{equation}
    \label{eq:boundaryLayer}
    -c_0^2\Delta S^+_\ell(\vec{\xi}) + \Delta^2 S^+_\ell(\vec{\xi}) + \gamma^{-1}\nabla\!\cdot\! \nabla\left[\vg{\vec{n}\cdot\underline{\underline{\mathcal{H}S^+_\ell}}\vec{n}}\right] = 0.
\end{equation}
Some solutions of this equation are given by solutions of
\begin{equation}
    \label{eq:boundaryLayer2}
    -c_0 S^+_\ell(\vec{\xi}) + \Delta S^+_\ell(\vec{\xi}) + \gamma^{-1}\vg{\vec{n}\cdot\underline{\underline{\mathcal{H}S^+_\ell}}\vec{n}}= 0,
\end{equation}
and using the assumption that $\underline{\underline{\nabla \vec{n}}} \equiv 0$ on the acoustic length scale \uz{and \eqref{eq:vectorCalculusIdentity}} this can be rewritten as
\begin{equation}
    \label{eq:boundaryLayer3}
    -c_0 S^+_\ell(\vec{\xi}) + \nabla \cdot \left[ (I + \gamma^{-1}\vec{n}\otimes\vec{n})\nabla S^+_\ell\right] = 0.
\end{equation}
From this we see that the boundary layer $S^+_\ell$ is governed by a reaction-diffusion equation with a transversally isotropic diffusion tensor $(I + \gamma^{-1}\vec{n}\otimes\vec{n})$.
Using the Vishik--Lyusternik method \cite{VishikLyusternik} we can assume the solution of \eqref{eq:scatteringProblem} has the form 
\begin{equation}
    \label{eq:vishikLyusternik}
    S^+(\vec{x}) = S^+_0(\vec{x}) + S^+_\ell(\vec{\xi}),
\end{equation}
where $S^+_0(\vec{x})$ is the solution of the Helmholtz equation for $\abs{\vec{x}}>1$.
Notice now that imposing the sound soft boundary conditions 
\begin{align}
    \label{eq:boundaryLayerConditions}
    \Delta S^+ + \gamma^{-1}\vec{n}\cdot \underline{\underline{\mathcal{H}}}S^+\vec{n} &= \Delta S^+_0 + \gamma^{-1}\vec{n}\cdot \underline{\underline{\mathcal{H}}}S^+_0\vec{n}\nonumber \\ 
    &+ \ell^{-2}\left(\Delta S^+_\ell + \gamma^{-1}\vec{n}\cdot \underline{\underline{\mathcal{H}}}S^+_\ell\vec{n}\right)\nonumber\\
    &= \Delta S^+_0 + \gamma^{-1}\vec{n}\cdot \underline{\underline{\mathcal{H}}}S^+_0\vec{n} + c_0 \ell^{-2} S^+_\ell\nonumber\\
    &= \Delta S^- + \gamma^{-1}\vec{n}\cdot \underline{\underline{\mathcal{H}}}S^- \vec{n},
\end{align}
therefore we can conclude that along the boundary of the circular obstacle the boundary layer $S^+_\ell$ has value
\begin{equation}
    \label{eq:boundaryLayerValue}
    S^+_\ell = \frac{\ell^2}{c_0}\left(\Delta S^- + \gamma^{-1}\vec{n}\cdot \underline{\underline{\mathcal{H}}}S^- \vec{n} - \Delta S^+_0 - \gamma^{-1}\vec{n}\cdot \underline{\underline{\mathcal{H}}}S^+_0\vec{n}\right).
\end{equation}
Since we know the value of the boundary layer $S^+_\ell$ on the perimeter of the circular obstacle we can rewrite the $S^+$ as being an $\mathcal{O}(\ell^2)$ perturbation of the solution of the following Helmholtz scattering problem
    \begin{alignat}{3}
        \label{eq:scatteringProblem2}
        -\omega^2S^+(\vec{x})-c_0^2\Delta S^+(\vec{x}) &= 0,\qquad && \abs{\vec{x}}>1, \nonumber \\
        S^+(\vec{x}) &= S^-(\vec{x})+\mathcal{O}(\ell^2),\qquad && \abs{\vec{x}}=1, \nonumber \\
        \abs{\partial_{\abs{\vec{x}}}S^+(\vec{x})-ikS^+(\vec{x})}&=\mathcal{O}(\abs{\vec{x}}^{-\frac{1}{2}}),\qquad && \abs{\vec{x}} \to \infty.
    \end{alignat}
\uz{
The solution of the previous equation can be expressed as a Mie series \cite{Moiola}, i.e.
\begin{equation}
    \label{eq:MieSeries}
    S^+(r,\theta) = - \sum_{j\in\mathbb{Z}} a_j \frac{H_j^{(1)}(kr)}{H_j^{(1)}(kR)}e^{ij\theta},
\end{equation}
where $(r,\theta)$ are the polar coordinates of $\vec{x}$, $R$ is the radius of the circular obstacle, $H_j^{(1)}$ are the Hankel functions of the first kind, $k$ is the wave number, and $a_j$ are the coefficients obtained by expanding the incoming plane wave $S^-$ in circular harmonics via the Jacobi--Anger formula \cite[eq.~(22)]{Moiola}, i.e.
\begin{equation}
    \label{eq:JacobiAnger}
    S^-(r,\theta) = e^{i\vec{k}\cdot \vec{x}} = \sum_{j\in\mathbb{Z}} a_j e^{ij\theta}, \qquad a_j = i^j e^{-ij\psi}J_j(kr),
\end{equation}
where $J_j$ is the Bessel function of the first kind and $\psi$ is the angle of incidence of the incoming plane wave.
}
Notice now that the wave number $k$ appearing in \eqref{eq:JacobiAnger} depends on the incoming plane wave and needs to satisfy the dispersion relation \eqref{eq:NematicQuadraticEquation}.
In particular, as discussed in Section \ref{sec:planeWaves}, the wave number $k$ is \uz{smaller} when the wave propagates orthogonal to the nematic director field. In regions where the incident plane wave is orthogonal to the nematic director field we can expect the amplitude of the reflected wave to be smaller, \uz{given the following asymptotic expansion of $S^+$ \cite[eq.~(38)]{Moiola}:
\begin{equation}
    S^+(r,\theta) = -\sqrt{\frac{2}{\pi kr}} \sum_{j\in \mathbb{Z}} a_j \frac{e^{i(kr-j\frac{\pi}{2})-\frac{\pi}{4}+j\theta}}{H_j^{(1)}(kR)}, \qquad r\to \infty.
\end{equation}
}
This is a striking difference between the behavior of a plane boundary, as discussed in Section \ref{eq:planeWaveReflection}, and the behavior of a circular obstacle. In fact while a \uz{change} of the amplitude of the reflected wave is expected only for impedance boundary conditions in the case of a plane boundary, in the case of a circular obstacle a reduction of the amplitude of the reflected wave is expected also for sound-soft boundary conditions.
This behavior is also observed in numerical simulations, as can be seen in Figure \ref{fig:scattering}.

\section{Conclusion}
In this work, we have considered the propagation of time-harmonic sound waves in a fluid governed by the Euler--Korteweg equation.
We derived two new equations describing these phenomena, which we name the Helmholtz--Korteweg and nematic Helmholtz--Korteweg equations.

Our analysis yielded new insights into the behavior of plane and evanescent waves and their dependence on both the nematic orientation and boundary conditions.
We also studied the reflection and transmission of plane waves at interfaces, considering different boundary conditions, and highlighted the unique behavior of waves in nematic-Korteweg fluids, including their anisotropy.
Numerical simulations confirmed that the speed of sound is highly dependent on the orientation of the nematic director field.
Finally, we explored the scattering of acoustic waves by circular obstacles, demonstrating that the nematic director field significantly alters the wave amplitude in a manner that is qualitatively distinct from the behavior at plane boundaries.

These results contribute to a deeper understanding of acoustic wave propagation in complex fluids, with potential applications in the design of devices with tunable acoustic properties.

\begin{acknowledgements}
This work was funded by the Engineering and Physical Sciences Research Council
[grant numbers EP/R029423/1 and EP/W026163/1],
and by the Donatio Universitatis Carolinae Chair
“Mathematical modelling of multicomponent systems”.
\vg{The authors would like to thank E.~Virga for helpful discussions and comments on the first draft of this manuscript.
The authors would like to thank also T.~van~Beeck and C.~Lehrenfeld for their help in understanding the correct boundary conditions for the Helmholtz--Korteweg equation.}
For the purpose of open access, the author has applied a CC BY public copyright licence to any author accepted manuscript arising from this submission.
No new data were generated or analysed during this work.
\end{acknowledgements}

\appendix
\uz{\section{A vector calculus identities}
We have used the vector calculus identity
\begin{equation}
    \label{eq:vectorCalculusIdentity}
    \nabla\cdot \Big((\vec{n}\otimes \vec{n})\nabla S \Big) = \vec{n}\cdot \underline{\underline{\mathcal{H}S}}\vec{n}
\end{equation}
several times in this work
This identity is most easily derived using tensor notation, i.e.
\begin{align}
    \nabla\cdot \Big((\vec{n}\otimes \vec{n})\nabla S \Big) &=  (\vec{n}\otimes \vec{n})_{ij,i} (\nabla S)_j + (n\otimes n)_{ij} (\nabla S)_{j,i}\nonumber\\
    &= \nabla S \cdot \nabla \cdot (\vec{n}\otimes \vec{n}) + (\vec{n}\otimes \vec{n})^T : \underline{\underline{\mathcal{H}S}}.
\end{align}
Using the facts that $\vec{n}\otimes \vec{n}$ and $\underline{\underline{\mathcal{H}S}}$ are symmetric and $\nabla\cdot \vec{n} = 0$ we can rewrite the previous equation as
\begin{align}
    \nabla\cdot \Big((\vec{n}\otimes \vec{n})\nabla S \Big) &= (\vec{n}\otimes \vec{n})_{ij} \underline{\underline{\mathcal{H} S}}_{\,ij}= \vec{n}\cdot \underline{\underline{\mathcal{H}S}}\vec{n}.
\end{align}
}
\bibliography{refs}

\begin{thebibliography}{33}%
\makeatletter
\providecommand \@ifxundefined [1]{%
 \@ifx{#1\undefined}
}%
\providecommand \@ifnum [1]{%
 \ifnum #1\expandafter \@firstoftwo
 \else \expandafter \@secondoftwo
 \fi
}%
\providecommand \@ifx [1]{%
 \ifx #1\expandafter \@firstoftwo
 \else \expandafter \@secondoftwo
 \fi
}%
\providecommand \natexlab [1]{#1}%
\providecommand \enquote  [1]{``#1''}%
\providecommand \bibnamefont  [1]{#1}%
\providecommand \bibfnamefont [1]{#1}%
\providecommand \citenamefont [1]{#1}%
\providecommand \href@noop [0]{\@secondoftwo}%
\providecommand \href [0]{\begingroup \@sanitize@url \@href}%
\providecommand \@href[1]{\@@startlink{#1}\@@href}%
\providecommand \@@href[1]{\endgroup#1\@@endlink}%
\providecommand \@sanitize@url [0]{\catcode `\\12\catcode `\$12\catcode `\&12\catcode `\#12\catcode `\^12\catcode `\_12\catcode `\%12\relax}%
\providecommand \@@startlink[1]{}%
\providecommand \@@endlink[0]{}%
\providecommand \url  [0]{\begingroup\@sanitize@url \@url }%
\providecommand \@url [1]{\endgroup\@href {#1}{\urlprefix }}%
\providecommand \urlprefix  [0]{URL }%
\providecommand \Eprint [0]{\href }%
\providecommand \doibase [0]{https://doi.org/}%
\providecommand \selectlanguage [0]{\@gobble}%
\providecommand \bibinfo  [0]{\@secondoftwo}%
\providecommand \bibfield  [0]{\@secondoftwo}%
\providecommand \translation [1]{[#1]}%
\providecommand \BibitemOpen [0]{}%
\providecommand \bibitemStop [0]{}%
\providecommand \bibitemNoStop [0]{.\EOS\space}%
\providecommand \EOS [0]{\spacefactor3000\relax}%
\providecommand \BibitemShut  [1]{\csname bibitem#1\endcsname}%
\let\auto@bib@innerbib\@empty
\bibitem [{\citenamefont {Korteweg}(1901)}]{Korteweg}%
  \BibitemOpen
  \bibfield  {author} {\bibinfo {author} {\bibfnamefont {D.~J.}\ \bibnamefont {Korteweg}},\ }\bibfield  {title} {\bibinfo {title} {Sur la forme que prennent les \'equations du mouvements des fluides si l'on tient compte des forces capillaires caus\'ees par des variations de densit\'e consid\'erables mais connues et sur la th\'eorie de la capillarit\'e dans l'hypoth\`ese d'une variation continue de la densit\'e},\ }\href@noop {} {\bibfield  {journal} {\bibinfo  {journal} {Archives N\'eerlandaises des Sciences Exactes et Naturelles}\ }\textbf {\bibinfo {volume} {6}},\ \bibinfo {pages} {1} (\bibinfo {year} {1901})}\BibitemShut {NoStop}%
\bibitem [{\citenamefont {Truesdell}\ \emph {et~al.}(2004)\citenamefont {Truesdell}, \citenamefont {Noll},\ and\ \citenamefont {Antman}}]{TruesdellNoll}%
  \BibitemOpen
  \bibfield  {author} {\bibinfo {author} {\bibfnamefont {C.}~\bibnamefont {Truesdell}}, \bibinfo {author} {\bibfnamefont {W.}~\bibnamefont {Noll}},\ and\ \bibinfo {author} {\bibfnamefont {S.~S.}\ \bibnamefont {Antman}},\ }\href@noop {} {\emph {\bibinfo {title} {The non-linear field theories of mechanics}}}\ (\bibinfo  {publisher} {Springer},\ \bibinfo {address} {Berlin},\ \bibinfo {year} {2004})\BibitemShut {NoStop}%
\bibitem [{\citenamefont {Lauro}(2008)}]{LauroMagma}%
  \BibitemOpen
  \bibfield  {author} {\bibinfo {author} {\bibfnamefont {G.}~\bibnamefont {Lauro}},\ }\bibfield  {title} {\bibinfo {title} {A note on a {Korteweg} fluid and the hydrodynamic form of the logarithmic {S}chrödinger equation},\ }\href {https://doi.org/10.1080/03091920801956957} {\bibfield  {journal} {\bibinfo  {journal} {Geophysical and Astrophysical Fluid Dynamics}\ }\textbf {\bibinfo {volume} {102}},\ \bibinfo {pages} {373} (\bibinfo {year} {2008})}\BibitemShut {NoStop}%
\bibitem [{\citenamefont {Benzoni-Gavage}(2013)}]{BenzoniPlanarWaves}%
  \BibitemOpen
  \bibfield  {author} {\bibinfo {author} {\bibfnamefont {S.}~\bibnamefont {Benzoni-Gavage}},\ }\bibfield  {title} {\bibinfo {title} {{Planar traveling waves in capillary fluids}},\ }\href {https://doi.org/10.57262/die/1360092830} {\bibfield  {journal} {\bibinfo  {journal} {Differential and Integral Equations}\ }\textbf {\bibinfo {volume} {26}},\ \bibinfo {pages} {439 } (\bibinfo {year} {2013})}\BibitemShut {NoStop}%
\bibitem [{\citenamefont {Virga}(2009)}]{VirgaNematoacoustics}%
  \BibitemOpen
  \bibfield  {author} {\bibinfo {author} {\bibfnamefont {E.~G.}\ \bibnamefont {Virga}},\ }\bibfield  {title} {\bibinfo {title} {Variational theory for nematoacoustics},\ }\href {https://link.aps.org/doi/10.1103/PhysRevE.80.031705} {\bibfield  {journal} {\bibinfo  {journal} {Physical Review E}\ }\textbf {\bibinfo {volume} {80}},\ \bibinfo {pages} {031705} (\bibinfo {year} {2009})}\BibitemShut {NoStop}%
\bibitem [{\citenamefont {Capriz}(1989)}]{Capriz}%
  \BibitemOpen
  \bibfield  {author} {\bibinfo {author} {\bibfnamefont {G.~G.}\ \bibnamefont {Capriz}},\ }\href@noop {} {\emph {\bibinfo {title} {Continua with microstructure}}},\ \bibinfo {series} {Springer Tracts in Natural Philosophy}, Vol.~\bibinfo {volume} {35}\ (\bibinfo  {publisher} {Springer-Verlag},\ \bibinfo {address} {New York},\ \bibinfo {year} {1989})\BibitemShut {NoStop}%
\bibitem [{\citenamefont {Toupin}(1962)}]{Toupin1}%
  \BibitemOpen
  \bibfield  {author} {\bibinfo {author} {\bibfnamefont {R.~A.}\ \bibnamefont {Toupin}},\ }\bibfield  {title} {\bibinfo {title} {Elastic materials with couple-stresses},\ }\href {https://doi.org/10.1007/BF00253945} {\bibfield  {journal} {\bibinfo  {journal} {Archive for Rational Mechanics and Analysis}\ }\textbf {\bibinfo {volume} {11}},\ \bibinfo {pages} {385} (\bibinfo {year} {1962})}\BibitemShut {NoStop}%
\bibitem [{\citenamefont {Toupin}(1964)}]{Toupin2}%
  \BibitemOpen
  \bibfield  {author} {\bibinfo {author} {\bibfnamefont {R.~A.}\ \bibnamefont {Toupin}},\ }\bibfield  {title} {\bibinfo {title} {Theories of elasticity with couple-stress},\ }\href {https://doi.org/10.1007/BF00253050} {\bibfield  {journal} {\bibinfo  {journal} {Archive for Rational Mechanics and Analysis}\ }\textbf {\bibinfo {volume} {17}},\ \bibinfo {pages} {85} (\bibinfo {year} {1964})}\BibitemShut {NoStop}%
\bibitem [{\citenamefont {Fried}\ and\ \citenamefont {Gurtin}(2006)}]{FriedGurtin}%
  \BibitemOpen
  \bibfield  {author} {\bibinfo {author} {\bibfnamefont {E.}~\bibnamefont {Fried}}\ and\ \bibinfo {author} {\bibfnamefont {M.~E.}\ \bibnamefont {Gurtin}},\ }\bibfield  {title} {\bibinfo {title} {Tractions, balances, and boundary conditions for nonsimple materials with application to liquid flow at small-length scales},\ }\href {https://doi.org/10.1007/s00205-006-0015-7} {\bibfield  {journal} {\bibinfo  {journal} {Archive for Rational Mechanics and Analysis}\ }\textbf {\bibinfo {volume} {182}},\ \bibinfo {pages} {513} (\bibinfo {year} {2006})}\BibitemShut {NoStop}%
\bibitem [{\citenamefont {Giovangigli}(2020)}]{Giovangigli}%
  \BibitemOpen
  \bibfield  {author} {\bibinfo {author} {\bibfnamefont {V.}~\bibnamefont {Giovangigli}},\ }\bibfield  {title} {\bibinfo {title} {Kinetic derivation of diffuse-interface fluid models},\ }\href {https://link.aps.org/doi/10.1103/PhysRevE.102.012110} {\bibfield  {journal} {\bibinfo  {journal} {Physical Review E}\ }\textbf {\bibinfo {volume} {102}} (\bibinfo {year} {2020})}\BibitemShut {NoStop}%
\bibitem [{\citenamefont {Dunn}\ and\ \citenamefont {Serrin}(1985)}]{DunnSerrin}%
  \BibitemOpen
  \bibfield  {author} {\bibinfo {author} {\bibfnamefont {J.~E.}\ \bibnamefont {Dunn}}\ and\ \bibinfo {author} {\bibfnamefont {J.}~\bibnamefont {Serrin}},\ }\bibfield  {title} {\bibinfo {title} {On the thermomechanics of interstitial working},\ }\href {https://doi.org/10.1007/BF00250907} {\bibfield  {journal} {\bibinfo  {journal} {Archive for Rational Mechanics and Analysis}\ }\textbf {\bibinfo {volume} {88}},\ \bibinfo {pages} {95} (\bibinfo {year} {1985})}\BibitemShut {NoStop}%
\bibitem [{\citenamefont {Anderson}\ \emph {et~al.}(1998)\citenamefont {Anderson}, \citenamefont {McFadden},\ and\ \citenamefont {Wheeler}}]{AndersonEtAll}%
  \BibitemOpen
  \bibfield  {author} {\bibinfo {author} {\bibfnamefont {D.~M.}\ \bibnamefont {Anderson}}, \bibinfo {author} {\bibfnamefont {G.~B.}\ \bibnamefont {McFadden}},\ and\ \bibinfo {author} {\bibfnamefont {A.~A.}\ \bibnamefont {Wheeler}},\ }\bibfield  {title} {\bibinfo {title} {Diffuse-interface methods in fluid mechanics},\ }\href {https://doi.org/10.1146/annurev.fluid.30.1.139} {\bibfield  {journal} {\bibinfo  {journal} {Annual Review of Fluid Mechanics}\ }\textbf {\bibinfo {volume} {30}},\ \bibinfo {pages} {139} (\bibinfo {year} {1998})}\BibitemShut {NoStop}%
\bibitem [{\citenamefont {Mehrabadi}\ \emph {et~al.}(2005)\citenamefont {Mehrabadi}, \citenamefont {Cowin},\ and\ \citenamefont {Massoudi}}]{MehrabadiEtAll}%
  \BibitemOpen
  \bibfield  {author} {\bibinfo {author} {\bibfnamefont {M.~M.}\ \bibnamefont {Mehrabadi}}, \bibinfo {author} {\bibfnamefont {S.~C.}\ \bibnamefont {Cowin}},\ and\ \bibinfo {author} {\bibfnamefont {M.}~\bibnamefont {Massoudi}},\ }\bibfield  {title} {\bibinfo {title} {Conservation laws and constitutive relations for density-gradient-dependent viscous fluids},\ }\href {https://doi.org/10.1007/s00161-004-0197-x} {\bibfield  {journal} {\bibinfo  {journal} {Continuum Mechanics and Thermodynamics}\ }\textbf {\bibinfo {volume} {17}},\ \bibinfo {pages} {183} (\bibinfo {year} {2005})}\BibitemShut {NoStop}%
\bibitem [{\citenamefont {Benzoni-Gavage}\ \emph {et~al.}(2005)\citenamefont {Benzoni-Gavage}, \citenamefont {Descombes}, \citenamefont {Jamet},\ and\ \citenamefont {Mazet}}]{BenzoniGavageHamiltonian}%
  \BibitemOpen
  \bibfield  {author} {\bibinfo {author} {\bibfnamefont {S.}~\bibnamefont {Benzoni-Gavage}}, \bibinfo {author} {\bibfnamefont {S.}~\bibnamefont {Descombes}}, \bibinfo {author} {\bibfnamefont {D.}~\bibnamefont {Jamet}},\ and\ \bibinfo {author} {\bibfnamefont {L.}~\bibnamefont {Mazet}},\ }\bibfield  {title} {\bibinfo {title} {Structure of {Korteweg} models and stability of diffuse interfaces},\ }\href {https://doi.org/10.4171/ifb/130} {\bibfield  {journal} {\bibinfo  {journal} {Interfaces and Free Boundaries}\ }\textbf {\bibinfo {volume} {7}},\ \bibinfo {pages} {371} (\bibinfo {year} {2005})}\BibitemShut {NoStop}%
\bibitem [{\citenamefont {Benzoni-Gavage}(2010)}]{BenzoniCIRM}%
  \BibitemOpen
  \bibfield  {author} {\bibinfo {author} {\bibfnamefont {S.}~\bibnamefont {Benzoni-Gavage}},\ }\href {https://math.univ-lyon1.fr/~benzoni/Levico.pdf} {\bibinfo {title} {Propagating phase boundaries and capillary fluids}} (\bibinfo {year} {2010}),\ \bibinfo {note} {notes from the international summer school on ``Mathematical Fluid Dynamics'', held at Levico Terme (Trento).}\BibitemShut {Stop}%
\bibitem [{\citenamefont {Benzoni-Gavage}\ \emph {et~al.}(2007)\citenamefont {Benzoni-Gavage}, \citenamefont {Danchin},\ and\ \citenamefont {Descombes}}]{BenzoniGavageWellposedness}%
  \BibitemOpen
  \bibfield  {author} {\bibinfo {author} {\bibfnamefont {S.}~\bibnamefont {Benzoni-Gavage}}, \bibinfo {author} {\bibfnamefont {R.}~\bibnamefont {Danchin}},\ and\ \bibinfo {author} {\bibfnamefont {S.}~\bibnamefont {Descombes}},\ }\bibfield  {title} {\bibinfo {title} {On the well-posedness for the {Euler-Korteweg} model in several space dimensions},\ }\href {http://www.jstor.org/stable/24902741} {\bibfield  {journal} {\bibinfo  {journal} {Indiana University Mathematics Journal}\ }\textbf {\bibinfo {volume} {56}},\ \bibinfo {pages} {1499} (\bibinfo {year} {2007})}\BibitemShut {NoStop}%
\bibitem [{\citenamefont {Murata}\ and\ \citenamefont {Shibata}(2020)}]{MurataShibata}%
  \BibitemOpen
  \bibfield  {author} {\bibinfo {author} {\bibfnamefont {M.}~\bibnamefont {Murata}}\ and\ \bibinfo {author} {\bibfnamefont {Y.}~\bibnamefont {Shibata}},\ }\bibfield  {title} {\bibinfo {title} {The global well-posedness for the compressible fluid model of {Korteweg} type},\ }\href {https://doi.org/10.1137/19M1282076} {\bibfield  {journal} {\bibinfo  {journal} {SIAM Journal on Mathematical Analysis}\ }\textbf {\bibinfo {volume} {52}},\ \bibinfo {pages} {6313} (\bibinfo {year} {2020})}\BibitemShut {NoStop}%
\bibitem [{\citenamefont {Tsuda}(2016)}]{Tsuda}%
  \BibitemOpen
  \bibfield  {author} {\bibinfo {author} {\bibfnamefont {K.}~\bibnamefont {Tsuda}},\ }\bibfield  {title} {\bibinfo {title} {On the existence and stability of time periodic solution to the compressible {Navier--Stokes} equation on the whole space},\ }\href {https://doi.org/10.1007/s00205-015-0902-x} {\bibfield  {journal} {\bibinfo  {journal} {Archive for Rational Mechanics and Analysis}\ }\textbf {\bibinfo {volume} {219}},\ \bibinfo {pages} {637} (\bibinfo {year} {2016})}\BibitemShut {NoStop}%
\bibitem [{\citenamefont {Giesselmann}\ \emph {et~al.}(2017)\citenamefont {Giesselmann}, \citenamefont {Lattanzio},\ and\ \citenamefont {Tzavaras}}]{TzavarasEtAll}%
  \BibitemOpen
  \bibfield  {author} {\bibinfo {author} {\bibfnamefont {J.}~\bibnamefont {Giesselmann}}, \bibinfo {author} {\bibfnamefont {C.}~\bibnamefont {Lattanzio}},\ and\ \bibinfo {author} {\bibfnamefont {A.~E.}\ \bibnamefont {Tzavaras}},\ }\bibfield  {title} {\bibinfo {title} {Relative energy for the {Korteweg} theory and related {H}amiltonian flows in gas dynamics},\ }\href {https://doi.org/10.1007/s00205-016-1063-2} {\bibfield  {journal} {\bibinfo  {journal} {Archive for Rational Mechanics and Analysis}\ }\textbf {\bibinfo {volume} {223}},\ \bibinfo {pages} {1427} (\bibinfo {year} {2017})}\BibitemShut {NoStop}%
\bibitem [{\citenamefont {Sonnet}\ and\ \citenamefont {Virga}(2012)}]{VirgaSonnet}%
  \BibitemOpen
  \bibfield  {author} {\bibinfo {author} {\bibfnamefont {A.~M.}\ \bibnamefont {Sonnet}}\ and\ \bibinfo {author} {\bibfnamefont {E.~G.}\ \bibnamefont {Virga}},\ }\href@noop {} {\emph {\bibinfo {title} {Dissipative ordered fluids : theories for liquid crystals}}}\ (\bibinfo  {publisher} {Springer},\ \bibinfo {address} {New York},\ \bibinfo {year} {2012})\BibitemShut {NoStop}%
\bibitem [{\citenamefont {Benzoni-Gavage}\ and\ \citenamefont {Chiron}(2018)}]{BenzoniGavageLongAsymptotics}%
  \BibitemOpen
  \bibfield  {author} {\bibinfo {author} {\bibfnamefont {S.}~\bibnamefont {Benzoni-Gavage}}\ and\ \bibinfo {author} {\bibfnamefont {D.}~\bibnamefont {Chiron}},\ }\bibfield  {title} {\bibinfo {title} {Long wave asymptotics for the {Euler--Korteweg} system},\ }\href {https://doi.org/10.4171/rmi/985} {\bibfield  {journal} {\bibinfo  {journal} {Revista Matemática Iberoamericana}\ }\textbf {\bibinfo {volume} {34}},\ \bibinfo {pages} {245} (\bibinfo {year} {2018})}\BibitemShut {NoStop}%
\bibitem [{\citenamefont {Dion}\ and\ \citenamefont {Jacob}(1977)}]{DionJacob}%
  \BibitemOpen
  \bibfield  {author} {\bibinfo {author} {\bibfnamefont {J.~L.}\ \bibnamefont {Dion}}\ and\ \bibinfo {author} {\bibfnamefont {A.~D.}\ \bibnamefont {Jacob}},\ }\bibfield  {title} {\bibinfo {title} {A new hypothesis on ultrasonic interaction with nematic liquid crystal},\ }\href {https://doi.org/10.1063/1.89753} {\bibfield  {journal} {\bibinfo  {journal} {Applied Physics Letters}\ }\textbf {\bibinfo {volume} {31}},\ \bibinfo {pages} {490} (\bibinfo {year} {1977})}\BibitemShut {NoStop}%
\bibitem [{\citenamefont {Dion}(1979)}]{Dion}%
  \BibitemOpen
  \bibfield  {author} {\bibinfo {author} {\bibfnamefont {J.~L.}\ \bibnamefont {Dion}},\ }\bibfield  {title} {\bibinfo {title} {The orienting action of ultrasound on liquid crystals related to the theorem of minimum entropy production},\ }\href {https://doi.org/10.1063/1.326175} {\bibfield  {journal} {\bibinfo  {journal} {Journal of Applied Physics}\ }\textbf {\bibinfo {volume} {50}},\ \bibinfo {pages} {2965} (\bibinfo {year} {1979})}\BibitemShut {NoStop}%
\bibitem [{\citenamefont {Mullen}\ \emph {et~al.}(1972)\citenamefont {Mullen}, \citenamefont {Lüthi},\ and\ \citenamefont {Stephen}}]{MullenLuti}%
  \BibitemOpen
  \bibfield  {author} {\bibinfo {author} {\bibfnamefont {M.~E.}\ \bibnamefont {Mullen}}, \bibinfo {author} {\bibfnamefont {B.}~\bibnamefont {Lüthi}},\ and\ \bibinfo {author} {\bibfnamefont {M.~J.}\ \bibnamefont {Stephen}},\ }\bibfield  {title} {\bibinfo {title} {Sound velocity in a nematic liquid crystal},\ }\href {https://doi.org/10.1103/PhysRevLett.28.799} {\bibfield  {journal} {\bibinfo  {journal} {Physical Review Letters}\ }\textbf {\bibinfo {volume} {28}},\ \bibinfo {pages} {799} (\bibinfo {year} {1972})}\BibitemShut {NoStop}%
\bibitem [{\citenamefont {Selinger}\ \emph {et~al.}(2002)\citenamefont {Selinger}, \citenamefont {Spector}, \citenamefont {Greanya}, \citenamefont {Weslowski}, \citenamefont {Shenoy},\ and\ \citenamefont {Shashidhar}}]{SelingerEtAll}%
  \BibitemOpen
  \bibfield  {author} {\bibinfo {author} {\bibfnamefont {J.~V.}\ \bibnamefont {Selinger}}, \bibinfo {author} {\bibfnamefont {M.~S.}\ \bibnamefont {Spector}}, \bibinfo {author} {\bibfnamefont {V.~A.}\ \bibnamefont {Greanya}}, \bibinfo {author} {\bibfnamefont {B.~T.}\ \bibnamefont {Weslowski}}, \bibinfo {author} {\bibfnamefont {D.~K.}\ \bibnamefont {Shenoy}},\ and\ \bibinfo {author} {\bibfnamefont {R.}~\bibnamefont {Shashidhar}},\ }\bibfield  {title} {\bibinfo {title} {Acoustic realignment of nematic liquid crystals},\ }\href {https://doi.org/10.1103/PhysRevE.66.051708} {\bibfield  {journal} {\bibinfo  {journal} {Physical Review E}\ }\textbf {\bibinfo {volume} {66}},\ \bibinfo {pages} {051708} (\bibinfo {year} {2002})}\BibitemShut {NoStop}%
\bibitem [{\citenamefont {Farrell}\ \emph {et~al.}(2023)\citenamefont {Farrell}, \citenamefont {Russo},\ and\ \citenamefont {Zerbinati}}]{FarrellEtAll}%
  \BibitemOpen
  \bibfield  {author} {\bibinfo {author} {\bibfnamefont {P.~E.}\ \bibnamefont {Farrell}}, \bibinfo {author} {\bibfnamefont {G.}~\bibnamefont {Russo}},\ and\ \bibinfo {author} {\bibfnamefont {U.}~\bibnamefont {Zerbinati}},\ }\href {https://arxiv.org/abs/2312.15210} {\bibinfo {title} {Kinetic derivation of a compressible {Leslie--Ericksen} equation for rarified calamitic gases}} (\bibinfo {year} {2023}),\ \Eprint {https://arxiv.org/abs/2312.15210} {arXiv:2312.15210 [math-ph]} \BibitemShut {NoStop}%
\bibitem [{Note1()}]{Note1}%
  \BibitemOpen
  \bibinfo {note} {Since the Helmholtz--Korteweg equation is fourth order, we employ $C^1$-conforming Argyris finite elements~\cite {Argyris} in Firedrake~\cite {firedrake}. The PDE analysis for these equations, and a numerical analysis of their discretisations, will be presented elsewhere.}\BibitemShut {Stop}%
\bibitem [{\citenamefont {Virga}(1994)}]{Virga}%
  \BibitemOpen
  \bibfield  {author} {\bibinfo {author} {\bibfnamefont {E.~G.}\ \bibnamefont {Virga}},\ }\href@noop {} {\emph {\bibinfo {title} {Variational theories for liquid crystals}}},\ \bibinfo {series} {Applied Mathematics and Mathematical Computation}, Vol.~\bibinfo {volume} {8}\ (\bibinfo  {publisher} {CRC Press, Taylor \& Francis Group},\ \bibinfo {year} {1994})\BibitemShut {NoStop}%
\bibitem [{\citenamefont {Vishik}\ and\ \citenamefont {Lyusternik}(1960)}]{VishikLyusternik}%
  \BibitemOpen
  \bibfield  {author} {\bibinfo {author} {\bibfnamefont {M.~I.}\ \bibnamefont {Vishik}}\ and\ \bibinfo {author} {\bibfnamefont {L.~A.}\ \bibnamefont {Lyusternik}},\ }\bibfield  {title} {\bibinfo {title} {The solution of some perturbation problems for matrices and selfadjoint or non-selfadjoint differential equations},\ }\href {https://10.1070/RM1960v015n03ABEH004092} {\bibfield  {journal} {\bibinfo  {journal} {Russian Mathematical Surveys}\ }\textbf {\bibinfo {volume} {15}},\ \bibinfo {pages} {1} (\bibinfo {year} {1960})}\BibitemShut {NoStop}%
\bibitem [{\citenamefont {Moiola}(2024)}]{Moiola}%
  \BibitemOpen
  \bibfield  {author} {\bibinfo {author} {\bibfnamefont {A.}~\bibnamefont {Moiola}},\ }\bibfield  {title} {\bibinfo {title} {Scattering of time-harmonic acoustic waves: Helmholtz equation, boundary integral equations and bem},\ }\href {https://mate.unipv.it/moiola/T/MNAPDE2024/MNAPDE2024.pdf} {\bibfield  {journal} {\bibinfo  {journal} {Lecture notes for the “Advanced numerical methods for PDEs” class, University of Pavia, Department of Mathematics}\ } (\bibinfo {year} {2024})}\BibitemShut {NoStop}%
\bibitem [{\citenamefont {Oskooi}\ \emph {et~al.}(2008)\citenamefont {Oskooi}, \citenamefont {Zhang}, \citenamefont {Avniel},\ and\ \citenamefont {Johnson}}]{oskooi2008}%
  \BibitemOpen
  \bibfield  {author} {\bibinfo {author} {\bibfnamefont {A.~F.}\ \bibnamefont {Oskooi}}, \bibinfo {author} {\bibfnamefont {L.}~\bibnamefont {Zhang}}, \bibinfo {author} {\bibfnamefont {Y.}~\bibnamefont {Avniel}},\ and\ \bibinfo {author} {\bibfnamefont {S.~G.}\ \bibnamefont {Johnson}},\ }\bibfield  {title} {\bibinfo {title} {The failure of perfectly matched layers, and towards their redemption by adiabatic absorbers},\ }\href {https://doi.org/10.1364/oe.16.011376} {\bibfield  {journal} {\bibinfo  {journal} {Optics Express}\ }\textbf {\bibinfo {volume} {16}},\ \bibinfo {pages} {11376} (\bibinfo {year} {2008})}\BibitemShut {NoStop}%
\bibitem [{\citenamefont {Argyris}\ \emph {et~al.}(1968)\citenamefont {Argyris}, \citenamefont {Fried},\ and\ \citenamefont {Scharpf}}]{Argyris}%
  \BibitemOpen
  \bibfield  {author} {\bibinfo {author} {\bibfnamefont {J.~H.}\ \bibnamefont {Argyris}}, \bibinfo {author} {\bibfnamefont {I.}~\bibnamefont {Fried}},\ and\ \bibinfo {author} {\bibfnamefont {D.~W.}\ \bibnamefont {Scharpf}},\ }\bibfield  {title} {\bibinfo {title} {The {TUBA} family of plate elements for the matrix displacement method},\ }\href {https://doi.org/10.1017/S000192400008489X} {\bibfield  {journal} {\bibinfo  {journal} {The Aeronautical Journal}\ }\textbf {\bibinfo {volume} {72}},\ \bibinfo {pages} {701} (\bibinfo {year} {1968})}\BibitemShut {NoStop}%
\bibitem [{\citenamefont {Ham}\ \emph {et~al.}(2023)\citenamefont {Ham}, \citenamefont {Kelly}, \citenamefont {Mitchell}, \citenamefont {Cotter}, \citenamefont {Kirby}, \citenamefont {Sagiyama}, \citenamefont {Bouziani}, \citenamefont {Vorderwuelbecke}, \citenamefont {Gregory}, \citenamefont {Betteridge}, \citenamefont {Shapero}, \citenamefont {Nixon-Hill}, \citenamefont {Ward}, \citenamefont {Farrell}, \citenamefont {Brubeck}, \citenamefont {Marsden}, \citenamefont {Gibson}, \citenamefont {Homolya}, \citenamefont {Sun}, \citenamefont {McRae}, \citenamefont {Luporini}, \citenamefont {Gregory}, \citenamefont {Lange}, \citenamefont {Funke}, \citenamefont {Rathgeber}, \citenamefont {Bercea},\ and\ \citenamefont {Markall}}]{firedrake}%
  \BibitemOpen
  \bibfield  {author} {\bibinfo {author} {\bibfnamefont {D.~A.}\ \bibnamefont {Ham}}, \bibinfo {author} {\bibfnamefont {P.~H.~J.}\ \bibnamefont {Kelly}}, \bibinfo {author} {\bibfnamefont {L.}~\bibnamefont {Mitchell}}, \bibinfo {author} {\bibfnamefont {C.~J.}\ \bibnamefont {Cotter}}, \bibinfo {author} {\bibfnamefont {R.~C.}\ \bibnamefont {Kirby}}, \bibinfo {author} {\bibfnamefont {K.}~\bibnamefont {Sagiyama}}, \bibinfo {author} {\bibfnamefont {N.}~\bibnamefont {Bouziani}}, \bibinfo {author} {\bibfnamefont {S.}~\bibnamefont {Vorderwuelbecke}}, \bibinfo {author} {\bibfnamefont {T.~J.}\ \bibnamefont {Gregory}}, \bibinfo {author} {\bibfnamefont {J.}~\bibnamefont {Betteridge}}, \bibinfo {author} {\bibfnamefont {D.~R.}\ \bibnamefont {Shapero}}, \bibinfo {author} {\bibfnamefont {R.~W.}\ \bibnamefont {Nixon-Hill}}, \bibinfo {author} {\bibfnamefont {C.~J.}\ \bibnamefont {Ward}}, \bibinfo {author} {\bibfnamefont {P.~E.}\ \bibnamefont {Farrell}}, \bibinfo {author} {\bibfnamefont {P.~D.}\ \bibnamefont {Brubeck}}, \bibinfo {author} {\bibfnamefont {I.}~\bibnamefont {Marsden}}, \bibinfo {author} {\bibfnamefont {T.~H.}\ \bibnamefont {Gibson}}, \bibinfo {author} {\bibfnamefont {M.}~\bibnamefont {Homolya}}, \bibinfo {author} {\bibfnamefont {T.}~\bibnamefont {Sun}}, \bibinfo {author} {\bibfnamefont {A.~T.~T.}\ \bibnamefont {McRae}}, \bibinfo {author} {\bibfnamefont {F.}~\bibnamefont {Luporini}}, \bibinfo {author} {\bibfnamefont {A.}~\bibnamefont {Gregory}}, \bibinfo {author} {\bibfnamefont {M.}~\bibnamefont {Lange}}, \bibinfo {author} {\bibfnamefont {S.~W.}\ \bibnamefont {Funke}}, \bibinfo {author} {\bibfnamefont {F.}~\bibnamefont {Rathgeber}}, \bibinfo {author} {\bibfnamefont {G.-T.}\ \bibnamefont {Bercea}},\ and\ \bibinfo {author} {\bibfnamefont {G.~R.}\ \bibnamefont {Markall}},\ }\href {https://10.25561/104839} {\emph {\bibinfo {title} {Firedrake User Manual}}},\ \bibinfo {organization} {Imperial College London and University of Oxford and Baylor University and University of
  Washington},\ \bibinfo {edition} {1st}\ ed. (\bibinfo {year} {2023})\BibitemShut {NoStop}%
\end{thebibliography}%
\end{document}